\documentclass[11pt,a4paper]{article}
\usepackage[T1]{fontenc}
\usepackage[utf8]{inputenc}
\usepackage[english]{babel}
\usepackage{authblk}
\usepackage{amsthm}
\usepackage{bbm}
\usepackage{color}
\usepackage[colorlinks=true,urlcolor=blue,linkcolor=blue]{hyperref}
\hypersetup{citecolor=blue}
\usepackage{graphicx}
\usepackage[bottom]{footmisc}
\usepackage{fullpage}
\usepackage{calc}
\usepackage{amsmath,amssymb,bm}
\usepackage{amsfonts}
\usepackage{mathrsfs}
\usepackage[nottoc]{tocbibind}
\usepackage{graphicx}
\graphicspath{ {images/} }
\usepackage{fancyhdr}
\usepackage{tabularx}
\usepackage{array}
\usepackage{lscape}
\usepackage{array,multirow,makecell}
\setcellgapes{1pt}
\makegapedcells
\usepackage[table]{xcolor}
\usepackage{float}
\usepackage{caption}
\usepackage{subcaption}
\usepackage{pifont}
\usepackage[ruled,vlined,linesnumbered,noresetcount]{algorithm2e}
\usepackage{booktabs}
\usepackage{titlesec}
\usepackage[section]{placeins}
\usepackage{enumitem}

\newcolumntype{R}[1]{>{\raggedleft\arraybackslash }b{#1}}
\newcolumntype{L}[1]{>{\raggedright\arraybackslash }b{#1}}
\newcolumntype{C}[1]{>{\centering\arraybackslash }b{#1}}
\newcommand\numberthis{\addtocounter{equation}{1}\tag{\theequation}}

\newtheorem{thm}{Theorem}
\newtheorem{prop}{Proposition}

\newtheorem{coro}{Corollary}
\newtheorem{rem}{Remark}
\newtheorem{ass}{Assumption}
\newtheorem{lem}{Lemma}

\makeatletter
\renewcommand\paragraph{\@startsection{paragraph}{4}{\z@}
  {-3.25ex \@plus -1ex \@minus -0.2ex}
  {2.25ex \@plus .25ex}
  {\normalfont\normalsize\bfseries}}
\renewcommand\subparagraph{\@startsection{subparagraph}{5}{\z@}
  {-3.25ex \@plus -1ex \@minus -0.2ex}
  {2.25ex \@plus .25ex}
  {\normalfont\normalsize\bfseries}}
\def\toclevel@paragraph{4}
\def\toclevel@paragraph{5}
\def\l@paragraph{\@dottedtocline{4}{7em}{4em}}
\def\l@subparagraph{\@dottedtocline{5}{7em}{4em}}
\makeatother

\title{Market Impact: Empirical Evidence, Theory and Practice}
\author[*,**]{Emilio Said}

\affil[*]{Department of Mathematics and Statistics, University of Montreal, Montreal, Canada}
\affil[**]{Quantitative Research and Development, Abu Dhabi Investment Authority (ADIA), Abu Dhabi, United Arab Emirates}

\begin{document}

\maketitle

\begin{abstract}
We propose a theory of the market impact of metaorders based on a coarse-grained approach where the microscopic details of supply and demand is replaced by a single parameter $\rho \in [0,+\infty]$ shaping the supply-demand equilibrium and the market impact process during the execution of the metaorder. Our model provides an unified explanation of most of the empirical observations that have been reported and establishes a strong connection between the excess volatility puzzle and the order-driven view of the markets through the square-root law. 
\end{abstract}


\section{Introduction}
\label{introduction}

Market impact can be defined as the difference between the actual price trajectory of the asset after the order is released to the market and the price trajectory that would have occurred if the order were never sent to the market. Unfortunately both price trajectories are not observable simultaneously and this makes that market impact is often viewed as the \textit{Heisenberg uncertainty principle of finance}. Considering its conceptual and practical importance market impact estimation and modelling has become one of the main topic in market microstructure. Market impact models have been intensively studied and used by practitioners in order to have pre-trade estimates of their expected trading cost and to optimize their execution strategy. These models also serve as post-trade tools to compute performance benchmarks and to evaluate trading results. From a more economic perspective market impact reflects the balance of supply and demand.

It is commonly accepted that although market information can come through many channels, it has been shown since \cite{grossman1980impossibility} \cite{grossman1989informational} that market impact is the main mechanism whereby information is conveyed to the market through trade execution. The market microstructure literature has empirically studied diverse types of market impact models, namely the market impact of \textit{single transactions}, \textit{aggregate transactions} and \textit{metaorders}. Although all these studies have reported such impact as a concave function of the transaction volume, they suggest different functional forms likely attributable to the different markets, time scales and the ways these studies have been conducted.

The traditional view in finance is that market impact is just a reflection of information and postulates that the functional form of market impact is the expression of how informed the agents are who trade with a given volume. Another view has emerged in the 1980s \cite{shiller1980stock} \cite{summers1986does} proposing to break away from the the original school of thought of rational market participants basing their trading decisions solely on the fluctuations of the fundamental value and replacing it by an order-driven theory where all trades -- informed, random, large or small -- whether initiated by rational or non-rational agents contribute to the volatility. In this new paradigm the main driver of price changes is not anymore the fundamental value but the order flow itself. To illustrate this \textit{order-driven view of markets} \cite{gabaix2021search} have recently developed a framework -- \textit{the inelastic market hypothesis} -- to theoretically and empirically analyze the fluctuations of the aggregate stock market and found that investing \$1 in the stock market increases the market’s aggregate value by about \$5. Based both on an empirical analysis of the long term price response to funds' order flow and on an equilibrium model of mandate-constrained investment firms, the authors provide an explanation of random movements of the stock market hard to link to fundamentals and at the origin of bursts of volatility in the absence of any news \cite{cutler1988moves} \cite{cornell2013moves}.

In this paper we will present a model for the market impact of metaorders replicating most of the empirical observations. Particularly, we will show that the impacted price trajectories are regularly varying functions\footnote{A measurable function $f : [0,+\infty) \rightarrow [0,+\infty)$ is \textit{regularly varying of index} $\rho \geq 0$ if for all $\lambda > 0$, $\lim\limits_{x \rightarrow +\infty}{\displaystyle\frac{f(\lambda x)}{f(x)}} = \lambda^{\rho}$ (see \cite{bingham_goldie_teugels_1987} Chapter 1 for an exhaustive overview).} of the metaorder size if, and only if the ratio between the average impact and the peak impact\footnote{The impact at the end of the metaorder.} stabilizes around a certain value between $0$ and $1$. Furthermore, we will see that the \textit{square-root law} is a direct consequence of the establishment of such an equilibrium and shed some light on the volatility factor that appears in its expression by showing how is related to the order-driven view of markets. We will also introduce \textit{the fair pricing point} of a metaorder and present some of its remarkable properties.

The paper is organized as follows: Section \ref{empirical evidence} presents the main empirical results that have been reported over the last years. Section \ref{theory} recalls some attempts that have been proposed to explain the square-root law. Section \ref{our contribution} provides an overview of our contribution. Sections \ref{model description}-\ref{averaging} introduce our model description and present our main findings. Section \ref{excess volatility and the order driven view of markets} sheds some light on the excess volatility puzzle. Section \ref{metaorders in practice} is a discussion of our results and their implications in practice. We conclude in Section \ref{conclusion}.

\section{Empirical Evidence}
\label{empirical evidence}

\subsection{Single Transactions Market Impact}
\label{single transactions market impact}

\cite{lillo2003master} studied the 1,000 largest stocks on the New York Stock Exchange in 1995-1998 and concluded that the price impact function of a single transaction is a power function of the transaction volume with the exponent between $0.1$ and $0.5$. \cite{hopman2007supply} came to the same conclusion by examining the Paris stock exchange from 1995 to 1999 and reported an average exponent of $0.4$. However \cite{potters2003more} found that a logarithmic function provides a better fit. Single transaction market impact is often referred to as \textit{price impact}.

\subsection{Aggregate Transactions Market Impact}
\label{aggregate transactions market impact}

\cite{gabaix2003theory} used the US transaction data over 15 minutes intervals to study the market impact of aggregate transactions i.e. the sum of signed transaction volumes over a given number of trades or times intervals and they found that the market impact function of those aggregate transactions is consistent with a power function of exponent $0.5$.

\subsection{Metaorders Market Impact}
\label{metaorders market impact}

Metaorders refer to large orders that are split into smaller pieces before being sent to the market to be executed. The study of the market impact of metaorders requires a different approach compared to individual or aggregate orders as it generates strong correlations in order flow through a sequence of incremental executions. From a microstructural point of view each trade can be seen as \textit{agressive} (taking liquidity) or \textit{passive} (providing liquidity) and can lead to an implicit order imbalance as the sum of agressive buy trades minus the sum of agressive sell trades. The resulting order flow can be generated by one or several market participants who share the same interests at the same time. In what follows we will refer to these order flows as metaorders even if they are not generated by a single market participant. Note that before the advent of the algorithmic trading era it was possible to identify and detect the direction of the aggressive order flow which explains why the conclusions reached by \cite{gabaix2003theory} for aggregate transactions are similar to the ones presented for metaorders.

In recent years many studies on different markets have been conducted to understand the influence of metaorders on the price formation process: \cite{almgren2005direct} \cite{moro2009market} \cite{toth2011anomalous} \cite{bershova2013non} \cite{bacry2015market} \cite{gomes2015market} \cite{said2017market} for the stock market, \cite{said2021market} for options market and \cite{donier2015million} for the bitcoin market. All of these studies agree that market impact is a two-phase process. A first characterized by a \textit{temporary} market impact -- concave and increasing with time --, followed by a relaxation -- convex and decreasing with time --, and giving rise to what is called the \textit{permanent} market impact. While most of studies agree on the properties of the temporary market impact, permanent market impact remains an object of controversy that will be discussed in Section \ref{different stories of permanent market impact}. A vast array of empirical studies have concluded that the impact of a metaorder scales roughly as a power function of its size. More precisely, in all of these cases\footnote{Except maybe \cite{zarinelli2015beyond} who have found that a logarithmic dependence fits better their data. In any case, all the functional forms that have been reported are regularly varying. Particularly, any logarithmic function is a regularly varying function with index $0$.} the impact at the end of the metaorder with participation rate $\mathcal{R}$\footnote{This theoretical form seems to be robust to the period chosen to estimate $\sigma$ and $\mathcal{R}$: Indifferently the different studies use the daily or the contemporaneous participation rate and volatility. An heuristic explanation, consistent only in the case $\delta = 1/2$, is given in \cite{bouchaud2018trades} (Chapter 12) based on the invariance of the liquidity ratio $\mathcal{L} := \displaystyle\frac{\sqrt{V}}{\sigma}$ (see Section \ref{excess volatility and the order driven view of markets}). In this paper, we will extend the strong relation highlighted by the invariance of the liquidity ratio between volume traded and volatility for any $\delta > 0$ (see Corollary \ref{corollary square root theorem equilibrium II}).} is well-described by a theoretical curve of the form $\propto \sigma \mathcal{R}^{\delta}$. We have summarized in the table below different measures of $\delta$. For the sake of clarity, we have only reported the studies that have explicitly suggested a market impact of the previous form.

\begin{center}
\begin{tabular}{l c c c}
\toprule[0.15 em]
Empirical Study & Market &
\multicolumn{1}{c}{$\mathcal{R}$} & \multicolumn{1}{c}{$\delta$} \\
\midrule[0.1 em]
\centering \cite{almgren2005direct} &
\centering Equity &
\centering daily &
\centering $0.6$ \tabularnewline
\centering \cite{toth2011anomalous} &
\centering Futures &
\centering daily &
\centering $0.5 - 0.6$ \tabularnewline
\centering \cite{bacry2015market} &
\centering Equity &
\centering daily &
\centering $0.53$ \tabularnewline
\centering \cite{brokmann2015slow} &
\centering Equity &
\centering execution time &
\centering $0.6$ \tabularnewline
\centering \cite{donier2015million} &
\centering Bitcoin &
\centering daily &
\centering $0.5$ \tabularnewline
\centering \cite{toth2016square} &
\centering Options &
\centering daily &
\centering $0.4 - 0.43$ \tabularnewline
\centering \cite{said2017market} &
\centering Equity &
\centering execution time &
\centering $0.5 - 0.6$ \tabularnewline
\centering \cite{bucci2018slow} &
\centering Equity &
\centering daily &
\centering $0.5$ \tabularnewline
\centering \cite{said2021market} &
\centering Options &
\centering daily &
\centering $0.53 - 0.56$ \tabularnewline
\bottomrule[0.15 em]
\end{tabular}
\captionof{table}{Values of $\delta$ that can be found in different empirical studies.}
\label{tab values of delta}
\end{center}

\noindent Table \ref{tab values of delta} seems to advocate in favor of $\delta \approx 0.5$ giving rise to what is now called the \textit{square-root law}. More surprisingly, the square-root law holds during the whole trajectory of metaorder and not only for the final execution price \cite{moro2009market} \cite{donier2015million} \cite{said2017market} \cite{said2021market} underlying the fact that the market does not anticipate the end of a metaorder.

\section{Theory}
\label{theory}

The first attempt to model metaorders market impact was initiated by \cite{kyle1985continuous} who developed a theory in a quadratic and Gaussian setup in which the total impact is a linear function of size in contradiction with empirical observations that show a good agreement with the square-root law. Since the mid-nineties several heuristic models have been proposed to account for the square-root law. One of them is based on the idea that the square-root behaviour reflects the risk of adverse selection and acts as a compensation for the market maker willing to take such a risk \cite{msci1997market} \cite{grinold2000active}. Based in this consideration, \cite{gabaix2003theory} proposed a theory, extended in \cite{gabaix2006institutional}, in which the liquidity provider has a mean variance utility function and is first-order risk averse. Since important efforts have been made to model non-linear market impact motivated by empirical studies and practitioners' needs. Mainly three approaches have gained traction over the last years.

\subsection{Propagator Models}
\label{propagator models}

Propagator models were introduced to reproduce the concave market impact shape of metaorders. In this setting each trade is assumed to have an instantaneous market impact which decreases according to a time-dependant decay kernel. Those models were studied in \cite{bouchaud2003fluctuations} \cite{almgren2005direct} \cite{obizhaeva2013optimal} before being generalized by \cite{gatheral2010no}. While these models yield fairly realistic results and are analytically tractable, they are purely phenomenological and may be inconsistent with empirical observation as underlined by their authors. Furthermore they do not provide a mechanism to explain impact.

\subsection{Limit Order Book Models}
\label{limit order book models}

This class of models has been developed in \cite{alfonsi2010optimal} \cite{alfonsi2010optimalexec} \cite{toth2011anomalous} \cite{alfonsi2012order} \cite{mastromatteo2014agent} and \cite{donier2015fully}. Limit order book models proposed an alternative theory based on a dynamical description of supply and demand in the order book. Although this approach offers a microstructural foundation of market impact, it suffers from ad hoc assumptions and approximations making it not realistic and practical. Furthermore it seems not to be consistent with empirical estimates of impact profiles, nor is it naturally consistent with the square-root law derived only in this setting in the case of metaorders executed at a constant rate. 

\subsection{Equilibrium Models}
\label{equilibrium models}

Another theory based on equilibrium considerations has been proposed by \cite{farmer2013efficiency}. In their framework, the authors assume a competitive equilibrium between liquidity providers and takers where the metaorders arrive sequentially with a volume distributed according to a power law. In addition they add two constraints during the metaorder execution to derive the square-root law: a martingale condition and a fair pricing hypothesis. The main objection to this model is the importance given to the metaorder size distribution which appears to be less universal than the square-root law. As an example, \cite{donier2015million} found that the square-root law holds in the Bitcoin market whereas the metaorder size distribution does not exactly follow the requirements of the Farmer's model. Another important limitation of the model is the consideration only of equally-sized execution strategies.

\section{Our Contribution}
\label{our contribution}

While the impact of single orders is non universal and highly sensitive to market microstructure and conditions, the impact of metaorders appears to be extremely robust against microstructural changes and always obeys to a square-root behaviour. The advent of high-frequency trading and regulatory changes have not even affected its validity. Considering this the striking question that immediately comes to mind is:
\begin{center}
    \emph{While the first trades have erratic price impacts, how the market impact of the entire metaorder smoothly turns into a stable square-root law?}\footnote{This question will be addressed in Section \ref{equilibrium}, see Theorems \ref{theorem equilibrium I} and \ref{theorem equilibrium II}.}
\end{center}
The universality of the square-root law suggests the existence of a \textit{coarse-graining} procedure which could explain and reproduce the main phenoma observed empirically while putting aside the microscopic details. Indeed, whatever the scale with we study and measure the effects of market impact we reach always the same conclusions. This line of reasoning appears in many situations in physics and it is also at the heart of the reaction-diffusion theory presented by \cite{donier2015fully}.

The model proposed here lies in this strand of research where a coarse-grained model is used to provide a simplified representation of the interactions between the liquidity takers and suppliers on the order book while keeping the main stylized facts. To this end, we have introduced a single emergent parameter $\rho \in [0,+\infty]$ encoding the microscopic specificities of the system and summarizing the intentions of the different agents. Heuristically, we can consider a group of traders submitting their orders to an algorithm in charge of dividing and executing incrementally the metaorder against a group of liquidity providers. The execution algorithm and the group of liquidity providers are subject to two very different types of incentives that affect their decision making. The algorithm wants to minimize its average market impact $\langle\mathcal{I}\rangle_n$ whereas the liquidity providers expect to maximize the peak impact $\mathcal{I}_n$ in compensation of their inventory risk\footnote{Equivalently, the algorithm is trying to minimize its average execution price and the liquidity providers are seeking to maximize the asset price in compensation of their inventory risk.} ($n$ is the number of child orders): An \textit{equilibrium} is reached when the ratio $\displaystyle\frac{\langle\mathcal{I}\rangle_n}{\mathcal{I}_n}$ stabilizes around $\displaystyle\frac{1}{1 + \rho} \in [0,1]$. In such an equilibrium, the market impact is expressed as a regularly varying function of the metaorder size, which means that $\mathcal{I}_n$ can be written $\mathcal{I}_n = (Q_1 + \dots + Q_n)^{\rho} \ell(Q_1 + \dots + Q_n)$ where $\ell(Q_1 + \dots + Q_n)$ is negligible\footnote{$\ell(Q_1 + \dots + Q_n) = o\left((Q_1 + \dots + Q_n)^{\lambda}\right)$ for every $\lambda > 0$ when $n \rightarrow +\infty$.} compared to $(Q_1 + \dots + Q_n)^{\rho}$. More surprisingly, the converse holds also. For instance, the validity of the  square-root law implies the existence of an equilibrium. This gives to our model a strong microstructural foundation since now the shape of the impact, mainly driven by $(Q_1 + \dots + Q_n)^{\rho}$, is directly connected to the supply-demand imbalance through the equilibrium ratio $\displaystyle\frac{1}{1 + \rho}$.\footnote{Note that $\rho = \displaystyle\frac{1}{2}$ which corresponds to the square-root law gives $\displaystyle\frac{1}{1 + \rho} = \displaystyle\frac{2}{3}$.} Perhaps even more importantly, many properties that have been empirically observed, such as the square-root law, are well explained in the context of these equilibriums. Our framework also addresses the \textit{excess volatility} puzzle and provides a description of the market in case of \textit{non-equilibrium} shedding some light on the the saw-tooth patterns that have been reported on several US securities on 19 July 2012.

\section{Model Description}
\label{model description}

Let $(\Omega, \mathcal{F}, \mathbb{P})$ a probability space and $(S_t)_{t \geq 0}$ a càdlàg process (right-continuous, left limits) being the stock price. We denote $\tau_1, \tau_2, \dots, \tau_n$ the execution times of a given metaorder of length $n$ (i.e. sliced in $n$ orders) and $Q_1, Q_2, \dots, Q_n \in [q_-,q_+]$, $0 < q_- \leq q_+ < +\infty$, the volumes mapped to these orders. Set $\varepsilon \in \{-1,1\}$ the sign of the metaorder and $S_{\tau_{1}^{-}} := \lim\limits_{\substack{t \rightarrow \tau_1 \\ t < \tau_1}}{S_t}$. Let equip the probability space with a filtration $(\mathcal{F}_n)_{n \geq 0}$ defined by $\mathcal{F}_0 :=  \sigma\left(\varepsilon, S_{\tau_1^-}\right)$ and for all $n \geq 1$, $\mathcal{F}_n := \sigma\left(\varepsilon, S_{\tau_1^-}, Q_1, \tau_1, \dots, Q_n, \tau_n\right)$. The definition of the filtration $(\mathcal{F}_n)_{n \geq 0}$ is extremely crucial and plays a central role in our modelization as it delimits the information the trader has in his possession and therefore conditions the results we can expect. Particularly, in our case the trader is perfectly agnostic of the market conditions and only has at his disposal the parameters of his execution scheduling i.e. the sign of the metaorder $\varepsilon$, the price at the beginning $S_{\tau_1^-}$, the execution times $\tau_1, \dots, \tau_n$ of the child orders and their corresponding sizes. Taking in consideration other parameters in our setting -- such as an accurate order flow imbalance estimator for instance -- could change the formulation of the problem and lead to different results. However, this is not anymore a market impact problem but it is more related to optimal execution which market impact estimation is only one of the sub problem that need to be addressed. Market impact is a market phenomenon that exists and affects all market participants, the way they want to deal with and address it in their optimal execution framework is up to them (see \cite{easley2015optimal} \cite{cartea2016incorporating} among other papers). \emph{Market impact, in essence, is always and everywhere a collective phenomenon that occurs in the direction of the market}. Rephrasing this in mathematical terms gives the assumptions \ref{assumption a.s. positivity} and \ref{assumption a.s. VWAP constraint} introduced in the next paragraph.

We assume also that for a given metaorder sliced in $n \geq 1$ orders, its final impact is always positive and its average execution price always lie between $S_{\tau_1^-}$ and $\mathbb{E}[S_{\tau_n}\,|\,\mathcal{F}_n]$ , when $n$ is large enough ($S_{\tau_1^-}$ and $\mathbb{E}[S_{\tau_n}\,|\,\mathcal{F}_n]$ being respectively the prices at the beginning and at the end of the metaorder). Hence, in what follows we will consider that

\begin{ass}[\textbf{a.s. positivity}]
\label{assumption a.s. positivity}
for any $n \geq 1$, $\mathbb{P}\left(\mathbb{E}\left[\varepsilon \left(S_{\tau_n} - S_{\tau_1^-}\right)\bigg|\mathcal{F}_n\right] > 0 \right) = 1$,
\end{ass}

\begin{ass}[\textbf{a.s. VWAP constraint}]
\label{assumption a.s. VWAP constraint}
$$ \mathbb{P}\left(\liminf\limits_{n \rightarrow +\infty}{\left\{\varepsilon S_{\tau_1^-} \leq \frac{\varepsilon\sum_{k=1}^{n}{Q_k\, \mathbb{E}[S_{\tau_k}\,|\,\mathcal{F}_k]}}{\sum_{k=1}^{n}{Q_k}} \leq \varepsilon\mathbb{E}[S_{\tau_n}\,|\,\mathcal{F}_n] \right\}}\right) = 1. $$
\end{ass}

\noindent Giving a metaorder we introduce $\mathcal{R}_n$ the \textit{friction} of the model such as
\begin{equation}
\label{definition R_n 1}
    \mathcal{R}_n \,\mathbb{E}[S_{\tau_n}\,|\,\mathcal{F}_n] + (1 - \mathcal{R}_n)\,S_{\tau_1^-} =  \frac{\displaystyle \sum_{k=1}^{n}{Q_k\, \mathbb{E}[S_{\tau_k}\,|\,\mathcal{F}_k]}}{\displaystyle \sum_{k=1}^{n}{Q_k}}, \quad n \geq 1.
\end{equation}
Note that Assumption \ref{assumption a.s. VWAP constraint} implies that a.s. $\mathcal{R}_n \in [0,1]$ when $n$ is large enough. Equivalently, (\ref{definition R_n 1}) may be written
\begin{equation}
\label{definition R_n 2}
    \mathcal{R}_n \,\mathbb{E}\left[\varepsilon \left(S_{\tau_n} - S_{\tau_1^-}\right)\bigg|\mathcal{F}_n\right] = \frac{\displaystyle \sum_{k=1}^{n}{Q_k\, \mathbb{E}\left[\varepsilon\left(S_{\tau_k} - S_{\tau_1^-}\right)\bigg|\mathcal{F}_k\right]}}{\displaystyle \sum_{k=1}^{n}{Q_k}}, \quad n \geq 1.
\end{equation}

\noindent Equation (\ref{definition R_n 2}) gives a straightforward interpretation of the term $\mathcal{R}_n$ by showing that for a metaorder of length $n$ it can be read as the ratio between the average impact and the peak impact i.e. the impact at the end of the metaorder. Setting the expected average impact
$$ \langle\mathcal{I}\rangle_n := \frac{\displaystyle \sum_{k=1}^{n}{Q_k\, \mathbb{E}\left[\varepsilon\left(S_{\tau_k} - S_{\tau_1^-}\right)\bigg|\mathcal{F}_k\right]}}{\displaystyle \sum_{k=1}^{n}{Q_k}} $$
and the expected impact
$$ \mathcal{I}_n := \mathbb{E}\left[\varepsilon \left(S_{\tau_n} - S_{\tau_1^-}\right)\bigg|\mathcal{F}_n\right], $$
we have for every $n \geq 1$,
$$ \mathcal{R}_n := \frac{\langle\mathcal{I}\rangle_n}{\mathcal{I}_n}. $$
The term \textit{friction} is justified by the fact that the liquidity takers aim to minimize $\langle \mathcal{I} \rangle_n$ whereas the liquidity providers want to maximize $\mathcal{I}_n$ at each transaction $n$. One has also to keep in mind that $\mathcal{I}_n$ may affect $\langle \mathcal{I} \rangle_n$ as it appears in its formulation. For example when $\mathcal{I}_n$ rises sharply, so does $\langle \mathcal{I} \rangle_n$.

\section{Notations}
\label{notations}

\begin{itemize}[label=$\blacktriangleright$]
    \item $(u_n)_{n \geq 1}$ and $(v_n)_{n \geq 1}$ being two sequences of positive real numbers, we will write $u_n = o(v_n)$ when $\lim\limits_{n \rightarrow +\infty}{\displaystyle\frac{u_n}{v_n}} = 0$. We will also write $u_n \sim_{+ \infty} v_n$ when $u_n - v_n = o(v_n)$ or equivalently $\lim\limits_{n \rightarrow +\infty}{\displaystyle\frac{u_n}{v_n}} = 1$.
    \item We will say that a real valued sequence (resp. a function) is said \textit{eventually} to have a certain property $P$ if $P(n)$ (resp. $P(x)$) is true for sufficiently large $n$ (resp. $x$).
\end{itemize}

\section{Equilibrium}
\label{equilibrium}

\subsection{\texorpdfstring{$\bm{\rho \in [0,+\infty)}$}{}}
\label{rho < infinity}

\begin{thm}[\textbf{metaorders market impact asymptotics}]
\label{theorem equilibrium I}
Let $\rho \in [0,+\infty)$. The following propositions are equivalent.
\begin{enumerate}[label=(\roman*)]
    \item $\lim\limits_{n \rightarrow +\infty}{\mathcal{R}_n} = \displaystyle\frac{1}{1 + \rho}$ and
    $\displaystyle\frac{\mathcal{R}_{n-1}}{\mathcal{R}_n} = 1 + o\left(\displaystyle\frac{Q_n}{Q_1 + \dots + Q_n}\right)$ as $n \rightarrow +\infty$.
    \item There exist $\eta \in \mathbb{R}$ and $\theta$ a bounded measurable function of a real variable such that for all $n \geq 1$,
    $$ \mathbb{E}\left[\varepsilon \left(S_{\tau_n} - S_{\tau_1^-}\right)\bigg|\mathcal{F}_n\right] = (Q_1 + \dots + Q_n)^{\rho}\exp\left(\eta + \int_{0}^{Q_1 + \dots + Q_n}{\frac{\theta(u)}{u}\,\mathrm{d}u}\right) $$
    where $\lim\limits_{x \rightarrow +\infty}{\theta(x)} = 0$.
    \item $\displaystyle\frac{\mathbb{E}\left[\varepsilon \left(S_{\tau_{n-1}} - S_{\tau_1^-}\right)\bigg|\mathcal{F}_{n-1}\right]}{\mathbb{E}\left[\varepsilon \left(S_{\tau_n} - S_{\tau_1^-}\right)\bigg|\mathcal{F}_n\right]} = 1 - \rho \frac{Q_n}{Q_1 + \dots + Q_n} + o\left(\frac{Q_n}{Q_1 + \dots + Q_n}\right)$ as $n \rightarrow +\infty$.
    \item $\left((Q_1 + \dots + Q_n)^{-\sigma} \mathbb{E}\left[\varepsilon \left(S_{\tau_n} - S_{\tau_1^-}\right)\bigg|\mathcal{F}_n\right]\right)_{n \geq 1}$ is eventually increasing for each $\sigma < \rho$ and  $\left((Q_1 + \dots + Q_n)^{-\tau} \mathbb{E}\left[\varepsilon \left(S_{\tau_n} - S_{\tau_1^-}\right)\bigg|\mathcal{F}_n\right]\right)_{n \geq 1}$ is eventually decreasing for all $\tau > \rho$.
\end{enumerate}
\end{thm}

\noindent Several properties of the market impact process $(\mathcal{I}_n)_{n \geq 1}$ appear implicitly in the propositions of Theorem \ref{theorem equilibrium I}.
\begin{enumerate}[label=$\star$]
    \item The second condition of proposition $(i)$ suggests that the process $(\mathcal{R}_n)_{n \geq 1}$ stabilizes slowly around $\displaystyle\frac{1}{1 + \rho}$. In contrast with Theorem \ref{theorem equilibrium II} we will see below how this additional asymptotic property on the speed of the establishment of the equilibrium provides particular properties to the impact process $(\mathcal{I}_n)_{n \geq 1}$.
    \item In $(ii)$ the impact $\mathcal{I}_n$ can be decomposed into two terms
    \begin{itemize}[label=$-$]
        \item $(Q_1 + \dots + Q_n)^{\rho}$ the effective pressure\footnote{To some extent its informational part, to not be confused with \textit{fundamental information} which suggests some knowledge of the fundamental price of the asset. It refers here rather to its contribution to the price formation process.} of the metaorder. This is the first-order term.
        \item $\exp\left(\eta + \displaystyle\int_{0}^{Q_1 + \dots + Q_n}{\frac{\theta(u)}{u}\,\mathrm{d}u}\right) = o\left((Q_1 + \dots + Q_n)^{\lambda}\right)$ for all $\lambda > 0$ when $n \rightarrow +\infty$. This term is negligible compared to the first term when the metaorder size is large enough and may be associated to the microscopic details of the order book. Note also that the market impact of a metaorder is only determined by its size $Q_1 + \dots + Q_n$ and does not show any explicit dependence on $\tau_1, \dots, \tau_n$.\footnote{One must be very cautious about such a statement. An equilibrium implies that things are happening under normal trading conditions. An as example, let us consider the case where the times between two consecutive trades are substantial. In that case, due to market conditions' changes it is not even clear that the process $(\mathcal{R}_n)_{n \geq 1}$ can converge. Implicitly, behind the establishment of an equilibrium there is a memory time and the idea that the timing and the participation rate during the execution of the metaorder are reasonable.}
    \end{itemize}
    \begin{rem}
    Keep in mind that in $\rho$ and $\eta$ are random variables and $\theta$ is a stochastic process. Hence, a priori, their behaviour can vary from a trajectory to another.   
    \end{rem}
    \item Another property is the asymptotic behaviour of the impact per share. Set $\mathcal{I}_0 := 0$ and $\delta_n := \mathcal{I}_n - \mathcal{I}_{n-1}$ the \textit{incremental impacts}, $n \geq 1$ and notice that
    $$ 1 - \frac{\mathcal{I}_{n-1}}{\mathcal{I}_n} = \frac{\delta_n}{\delta_1 + \dots + \delta_n}. $$
    Hence when $\rho > 0$, $(iii)$ may be rewritten as
    \begin{equation}
    \label{structure relative weights between impacts and sizes}
        \frac{\delta_n}{\delta_1 + \dots + \delta_n} \sim_{+\infty} \rho \frac{Q_n}{Q_1 + \dots + Q_n}
    \end{equation}
    or
    \begin{equation}
    \label{structure impact per share}
        \frac{\delta_n}{Q_n} \sim_{+\infty} \rho \frac{\delta_1 + \dots + \delta_n}{Q_1 + \dots + Q_n}.
    \end{equation}
    (\ref{structure relative weights between impacts and sizes}) draws an asymptotic linear relation between the relative weight of the $n^{\mathrm{th}}$ order over the size and the impact of the metaorder. It underlines also the mechanism of how trades impact prices in our model and can be read as the liquidity providers' attempts to forecast the amount of information contained in the trades in order to adjust the price up or down accordingly. This is more or less the same story that can be found in \cite{kyle1985continuous}. (\ref{structure impact per share}) gives the asymptotic behaviour of the impact per share of the $n^{\mathrm{th}}$ trade in contrast to the impact per share of the ongoing metaorder. Clearly the impact per share is amplified when $\rho > 1$ and mitigated when $\rho < 1$ leading asymptotically to a convex / concave shaped profile of the market impact process (see also Corollary \ref{corollary embedding concave/convex market impact function theorem equilibrium II}).
    \item From $(iv)$ the market impact $(\mathcal{I}_n)_{n \geq 1}$ is eventually increasing when $\rho > 0$. Hence, as expected during the execution of a buy (resp. sell) metaorder prices are going up (resp. down).
\end{enumerate}

\noindent Theorem \ref{theorem equilibrium I} can be weakened to give the following result.

\begin{thm}[\textbf{metaorders market impact asymptotics generalization}]
\label{theorem equilibrium II}
Let $\rho \in [0,+\infty)$. The following propositions are equivalent.
\begin{enumerate}[label=(\roman*)]
    \item $\lim\limits_{n \rightarrow +\infty}{\mathcal{R}_n} = \displaystyle\frac{1}{1 + \rho}$.
    \item There exist two bounded measurable functions $\eta$, $\theta$ of a real variable such that for all $n \geq 1$,
    $$ \mathbb{E}\left[\varepsilon \left(S_{\tau_n} - S_{\tau_1^-}\right)\bigg|\mathcal{F}_n\right] = (Q_1 + \dots + Q_n)^{\rho}\exp\left(\eta(Q_1 + \dots + Q_n) + \int_{0}^{Q_1 + \dots + Q_n}{\frac{\theta(u)}{u}\,\mathrm{d}u}\right) $$
    where $\lim\limits_{x \rightarrow +\infty}{\eta(x)} = \kappa \in \mathbb{R}$ and $\lim\limits_{x \rightarrow +\infty}{\theta(x)} = 0$.
    \item There exists a sequence of positive real numbers $(\alpha_n)_{n \geq 1}$ such that 
    $$ \mathbb{E}\left[\varepsilon \left(S_{\tau_n} - S_{\tau_1^-}\right)\bigg|\mathcal{F}_n\right] \sim_{+\infty} \alpha_n $$ 
    and 
    $$ \frac{\alpha_{n-1}}{\alpha_n} = 1 - \rho \frac{Q_n}{Q_1 + \dots + Q_n} + o\left(\frac{Q_n}{Q_1 + \dots + Q_n}\right) $$ 
    as $n \rightarrow +\infty$.
    \item There exist $\zeta \in \mathbb{R}$ and $\chi$ a bounded measurable function of a real variable such that for all $n \geq 1$,
    $$ \frac{\displaystyle \sum_{k=1}^{n}{Q_k\, \mathbb{E}\left[\varepsilon\left(S_{\tau_k} - S_{\tau_1^-}\right)\bigg|\mathcal{F}_k\right]}}{\displaystyle \sum_{k=1}^{n}{Q_k}} = (Q_1 + \dots + Q_n)^{\rho}\exp\left(\zeta + \int_{0}^{Q_1 + \dots + Q_n}{\frac{\chi(u)}{u}\,\mathrm{d}u}\right) $$
    where $\lim\limits_{x \rightarrow +\infty}{\chi(x)} = 0$.
\end{enumerate}
\end{thm}

\noindent Hence Theorem \ref{theorem equilibrium II} is an extension of Theorem \ref{theorem equilibrium I} with less restrictive conditions. Some comments are in order to put in perspective these two theorems.
\begin{enumerate}[label=$\star$]
    \item The extra term characterizing the speed of convergence to reach equilibrium has been dropped. However, by combining $(ii)$ and $(iv)$ $\mathcal{R}_n$ can be now expressed as a \textit{slowly varying function}\footnote{A measurable function $\ell : [0,+\infty) \rightarrow [0,+\infty)$ is \textit{slowly varying} if for all $\lambda > 0$, $\lim\limits_{x \rightarrow +\infty}{\displaystyle\frac{\ell(\lambda x)}{\ell(x)}} = 1$.} of the metaorder size (see Corollary \ref{corollary slowly varying function theorem equilibrium II}). This reinforces the idea that once reached this equilibrium is quite stable.
    \item The constant $\eta$ in Theorem \ref{theorem equilibrium I} $(ii)$ is now replaced by a convergent function.
    \item The sequence $(\mathcal{I}_n)_{n \geq 1}$ is now equivalent to another sequence $(\alpha_n)_{n \geq 1}$ having all the \textit{good properties} presented at the end of Theorem \ref{theorem equilibrium I}.
\end{enumerate}

\begin{coro}[\textbf{stable equilibrium}]
\label{corollary slowly varying function theorem equilibrium II}
Assume that the conditions of Theorem \ref{theorem equilibrium II} hold. Then there exists a slowly varying function $\ell : [0,+\infty) \rightarrow [0,1]$ such that for every $n \geq 1$, $\mathcal{R}_n = \ell(Q_1 + \dots + Q_n)$.
\end{coro}

\begin{rem}
Without loss of generality the slowly varying function $\ell$ in Corollary \ref{corollary slowly varying function theorem equilibrium II} can be taken $C^{\infty}$ (Proposition 1.3.4 in \cite{bingham_goldie_teugels_1987}).
\end{rem}

\noindent The fact that $\mathcal{R}_n$ can be expressed as a smooth slowly varying function of the metaorder size supports also the idea that the market impact of metaorders is an aggregate low frequency phenomena in contrast with the activity of opportunistic traders acting at shorter time scales with smaller order sizes.

\begin{coro}[\textbf{log impact}]
\label{corollary log equivalent theorem equilibrium II}
Assume that the conditions of Theorem \ref{theorem equilibrium II} hold. If $\rho > 0$, then
$$ \log\mathbb{E}\left[\varepsilon \left(S_{\tau_n} - S_{\tau_1^-}\right)\bigg|\mathcal{F}_n\right] \sim_{+\infty}  \rho \log\left(Q_1 + \dots + Q_n\right). $$
\end{coro}

\noindent We have seen that the market impact process is given by $\mathcal{I}_n = (Q_1 + \dots + Q_n)^{\rho} \ell(Q_1 + \dots + Q_n)$ where $\ell(Q_1 + \dots + Q_n)$ is a slowly varying second order term. Sometimes it is more convenient to express market impact as a percentage of the total volume of trade in the market to put things in perspective. Define $(V_n)_{n \geq 1}$ as the positive volumes traded in the market during the execution of the metaorder such as $V_1$ is the total volume traded between $\tau_1^{-}$ and $\tau_1$ and $V_n$ the volume exchanged between $\tau_{n-1}$ and $\tau_n$ for each $n \geq 2$. Hence for every $n \geq 1$, $Q_n \leq V_n$.

\begin{coro}[\textbf{market impact formula}]
\label{corollary square root theorem equilibrium II}
Assume that the conditions of Theorem \ref{theorem equilibrium II} hold and $\left(\displaystyle\frac{Q_1 + \dots + Q_n}{V_1 + \dots + V_n}\right)_{n \geq 1}$ converges to $\displaystyle\frac{Q}{V} \in (0,1]$. Set for all $x \geq 0$,
$$ f(x) := x^{\rho} \exp\left(\eta(x) + \int_{0}^{x}{\frac{\theta(u)}{u}\mathrm{d}u}\right) $$
and for every $n \geq 1$,
\begin{equation}
\label{endogenous volatility}
    \hat{\sigma}_n := f(V_1 + \dots + V_n).
\end{equation}
Then
\begin{equation}
\label{trajectorial square-root law}
    \mathbb{E}\left[\varepsilon \left(S_{\tau_n} - S_{\tau_1^-}\right)\bigg|\mathcal{F}_n\right] \sim_{+\infty} \hat{\sigma}_n \left(\frac{Q}{V}\right)^{\rho}.
\end{equation}
\end{coro}

\begin{rem}
Note that the sequence $(V_n)_{n \geq 1}$ could have been taken such as 
$$ \frac{Q}{V} := \lim\limits_{n \rightarrow +\infty}{\displaystyle\frac{Q_1 + \dots + Q_n}{V_1 + \dots + V_n}} $$ 
is the daily participation rate instead of the participation rate during the execution of the metaorder. 
\end{rem}

\begin{coro}[\textbf{concave / convex profiles}]
\label{corollary embedding concave/convex market impact function theorem equilibrium II}
Assume that the conditions of Theorem \ref{theorem equilibrium II} hold and $\lim\limits_{x \rightarrow +\infty}{f(x)} = +\infty$\footnote{When $\rho > 0$, it is always the case since $\log f(x) \sim_{+\infty} \rho \log x$.} with $f(x) := x^{\rho} \exp\left(\eta(x) + \displaystyle\int_{0}^{x}{\frac{\theta(u)}{u}\mathrm{d}u}\right)$ for every $x \geq 0$.
\begin{itemize}
    \item If $f$ is eventually concave, then $\rho \leq 1$.
    \item If $f$ is eventually convex, then $\rho \geq 1$.
\end{itemize}
\end{coro}

\subsection{\texorpdfstring{$\bm{\rho = +\infty}$}{}}
\label{rho = infinity}

Two extreme cases are $\rho = 0$ and $\rho = +\infty$. The case $\rho = 0$ has already been addressed in Section \ref{rho < infinity} and arises when $\langle \mathcal{I}_n \rangle_n \sim_{+\infty} \mathcal{I}_n$. In this case, by setting $\mathcal{R}_{\infty} := \lim\limits_{n \rightarrow +\infty}{\mathcal{R}_n} = \displaystyle\frac{1}{1 + \rho}$, we have $\mathcal{R}_{\infty} = 1$ in favor of the liquidity takers as the impact per share decreases during the execution of the metaorder. The case $\rho = +\infty$ or equivalently $\mathcal{R}_{\infty} = 0$ corresponds to $\langle \mathcal{I} \rangle_n = o\left(\mathcal{I}_n\right)$.

\begin{prop}[$\bm{\rho = +\infty}$]
\label{proposition rho infinity theorem equilibrium II}
If there exists a sequence of positive real numbers $(\alpha_n)_{\geq 1}$ such that
$$ \mathbb{E}\left[\varepsilon \left(S_{\tau_n} - S_{\tau_1^-}\right)\bigg|\mathcal{F}_n\right] \sim_{+\infty} \alpha_n \enspace\text{and}\enspace \lim\limits_{n \rightarrow +\infty}{\frac{Q_1 + \dots + Q_n}{Q_n}\left(1 - \frac{\alpha_{n-1}}{\alpha_n}\right)} = +\infty, $$
then $\lim\limits_{n \rightarrow +\infty}{\mathcal{R}_n} = 0$.
\end{prop}

\noindent In that situation the impact per share increases sharply and diverges in favor of the liquidity providers. Thus $\mathcal{R}_n$ can also be interpreted as a performance measure of the execution of the metaorder.

\section{Non-equilibrium}
\label{non-equilibrium}

We have seen in Section \ref{equilibrium} how the convergence of $(\mathcal{R}_n)_{n \geq 1}$ shapes the behaviour of the market impact process $(\mathcal{I}_n)_{n \geq 1}$. Now we want to investigate the case when such a convergence does not exist. In that situation, the sequence $(\mathcal{R}_n)_{n \geq 1}$ is still bounded and its set of limit points $\mathcal{L} \subset [0,1]$ has at least two elements which means that there is not a consensus among the liquidity providers. To illustrate this, let us consider $\mathcal{L}$ has exactly two elements $\displaystyle\frac{1}{1 + \rho_1}$ and $\displaystyle\frac{1}{1 + \rho_2}$ representing the beliefs of two distinct groups of liquidity providers. Without loss of generality $\rho_1$ and $\rho_2$ can be taken such as $\rho_1 < \rho_2$. Naturally, the liquidity takers in charge of the execution of the metaorder will start to trade with the first group until their maximum inventory risk has been reached and then with the second group. These non-equilibrium situations can arise when several liquidity providers who do not share the same belief about the future prices are in competition or when the market conditions evolve quickly during the metaorder's execution forcing the liquidity providers to review and adjust their inventory risk and quotes. These events are more likely when the metaorders last several days or weeks. Define $\rho_-$, $\rho_+ \in [0,+\infty]$ such that
$$ \frac{1}{1 + \rho_+} := \liminf\limits_{n \rightarrow +\infty}{\mathcal{R}_n} $$
and
$$ \frac{1}{1 + \rho_-} := \limsup\limits_{n \rightarrow +\infty}{\mathcal{R}_n}. $$
Hence $\mathcal{L} \subset \left[\displaystyle\frac{1}{1 + \rho_+}, \displaystyle\frac{1}{1 + \rho_-}\right]$, particularly
\begin{thm}[\textbf{non-equilibrium}]
\label{theorem set of limit points}
If $\left((Q_1 + \dots + Q_n) \mathbb{E}\left[\varepsilon \left(S_{\tau_n} - S_{\tau_1^-}\right)\bigg|\mathcal{F}_n\right]\right)_{n \geq 1}$ is eventually non-decreasing, then $\mathcal{L} = \left[\displaystyle\frac{1}{1 + \rho_+}, \displaystyle\frac{1}{1 + \rho_-}\right]$.
\end{thm}

\begin{figure}[H]
\centering
\includegraphics[scale=0.8]{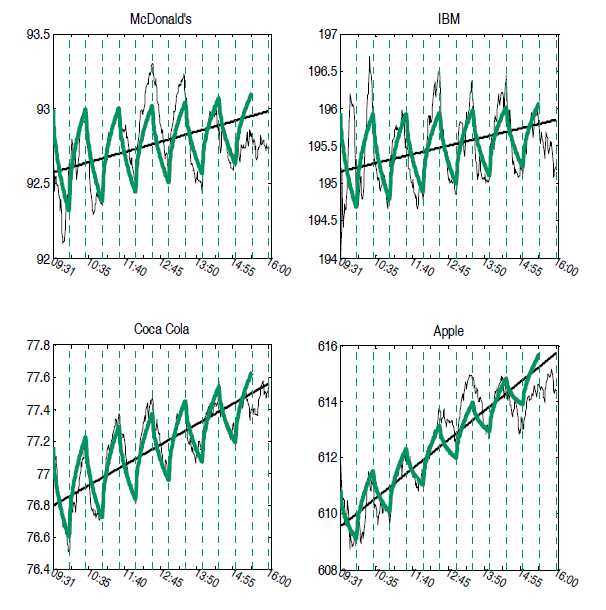}
\caption{Sawtooth patterns on Coca-Cola, Mcdonald's, IBM and Apple on 19 July 2012 from \cite{lehalle2012does}. The black line is the price, the green one is a model fitted by the authors.}
\label{Figure sawtooth patterns}
\end{figure}

\noindent In the non-equilibrium case, $(\mathcal{R}_n)_{n \geq 1}$ oscillates between $\displaystyle\frac{1}{1 + \rho_+}$ and $\displaystyle\frac{1}{1 + \rho_-}$ inducing similar swings on the price trajectory. These strange market impact effects (see Figure \ref{Figure sawtooth patterns}) have already been observed on four large cap US stocks (Coca-Cola, McDonald's, IBM and Apple) on 19 July 2012 \cite{lehalle2012does}.

\section{Averaging}
\label{averaging}

The results presented in Sections \ref{equilibrium} and \ref{non-equilibrium} are pathwise. However in practice, market impact is given on average over several metaorder's price trajectories. Set $\mathscr{C} := \left\{\left(\mathcal{R}_n\right)_{n \geq 1} \text{converges}\right\}$ and note that the random variables $\mathcal{R}_{\infty} := \lim\limits_{n \rightarrow +\infty}{\mathcal{R}_n}$ and $\rho$ are only well-defined on $\mathscr{C}$ and satisfy $\mathcal{R}_{\infty} = \displaystyle\frac{1}{1 + \rho}$. Note also that the set $\mathscr{C}$ may be written
$$ \mathscr{C} = \bigcap\limits_{\varepsilon \in \mathbb{Q}_+^*}\bigcup\limits_{N \geq 1}\bigcap\limits_{n \geq N}\bigcap\limits_{m \geq N}{\left\{\,|\mathcal{R}_n - \mathcal{R}_m| < \varepsilon\right\}} $$
and therefore $\mathscr{C} \in \mathcal{F}$. Suppose $\mathbb{P}(\mathscr{C}) > 0$ and define for all $A \in \mathcal{F}$,  $\mathbb{P}^{\mathscr{C}}(A) := \mathbb{P}\left(A\,|\,\mathscr{C}\right) = \displaystyle\frac{\mathbb{P}(A \cap \mathscr{C})}{\mathbb{P}(\mathscr{C})}$. In practice, $\mathscr{C}$ corresponds to all the possible configurations where the investors and market makers do not go bankrupt and act rationally. By Lebesgue's dominated convergence theorem
\begin{equation}
\label{average vwap relaxation}
    \lim\limits_{n \rightarrow +\infty}{\mathbb{E}^{\mathscr{C}}\left[\mathcal{R}_n\right]} = \mathbb{E}^{\mathscr{C}}\left[\frac{1}{1 + \rho}\right].
\end{equation}
Consider $(\hat{\sigma}_n)_{n \geq 1}$ and the random variable $\displaystyle\frac{Q}{V} \in (0,1]$ as defined in (\ref{endogenous volatility}) and (\ref{trajectorial square-root law}). Suppose also that $\rho$ and $\displaystyle\frac{Q}{V}$ are independent\footnote{The independence of the random variables $\rho$ and $\displaystyle\frac{Q}{V}$ can be interpreted as the liquidity providers' inability to detect the properties of the metaorders such as its size, starting and ending time.}. Again, a straightforward application of the Lebesgue's dominated convergence theorem leads to
\begin{equation}
\label{average normalized market impact function}
    \lim\limits_{n \rightarrow +\infty}{\mathbb{E}^{\mathscr{C}}\left[\frac{\mathcal{I}_n}{\hat{\sigma}_n}\,\Bigg|\,\frac{Q}{V}\right]} = \psi\left(\frac{Q}{V}\right)
\end{equation}
where the \textit{average normalized market impact function} $\psi : (0,1] \rightarrow (0,1]$ is given by
\begin{equation}
\label{average normalized market impact function definition}
    \psi(x) := \mathbb{E}^{\mathscr{C}}\left[x^{\rho}\right].
\end{equation}

\begin{prop}[\textbf{some properties of the function $\psi$}]
\label{proposition average normalized market impact function}
$\psi$ is continuous non-decreasing and $\lim\limits_{x \rightarrow 0^+}{\psi(x)} = \mathbb{P}^{\mathscr{C}}\left(\rho = 0\right)$. Furthermore
\begin{itemize}
    \item If $\rho \leq 1$ (resp. $\geq 1$) $\mathbb{P}^{\mathscr{C}}-$a.s., then $\psi$ is concave (resp. convex).
    \item If $\mathbb{P}^{\mathscr{C}}(\rho > 1) = 1$, then $\lim\limits_{x \rightarrow 0^+}{\displaystyle\frac{\psi(x)}{x}} = 0$.
    \item If $\mathbb{P}^{\mathscr{C}}(\rho = 0) = 0$ and $\mathbb{P}^{\mathscr{C}}(0 < \rho < 1) > 0$, then $\lim\limits_{x \rightarrow 0^+}{\displaystyle\frac{\psi(x)}{x}} = +\infty$.
    \item If $\rho$ is $\mathbb{P}^{\mathscr{C}}-$integrable, then $\psi$ is differentiable and for $x \in (0,1]$, $\psi'(x) = \mathbb{E}^{\mathscr{C}}[\rho x^{\rho-1}]$.
\end{itemize}
\end{prop}

\noindent Equations (\ref{average vwap relaxation}), (\ref{average normalized market impact function}) and (\ref{average normalized market impact function definition}) underline the fact that the probability distribution of the random variable $\rho$ is the key component of market impact studies.

\subsection{Special cases}
\label{special cases}

\begin{minipage}{\textwidth}
\centering
\begin{tabular}{l c c c}
\toprule[0.15 em]
Distribution\footnotemark & 
\multicolumn{1}{c}{$\mathbb{E}^{\mathscr{C}}[\rho]$} & \multicolumn{1}{c}{$\mathbb{E}^{\mathscr{C}}\left[\displaystyle\frac{1}{1 + \rho}\right]$} &
\multicolumn{1}{c}{$\psi(x)$} \\
\midrule[0.1 em]
\centering $\mathcal{D}\left(\frac{1}{2}\right)$ & 
\centering $1/2$ & 
\centering $2/3$ & 
\centering $\sqrt{x}$ \tabularnewline
\centering $\mathcal{U}\left(\left[0,1\right]\right)$ & 
\centering $1/2$ & 
\centering $\log 2$ & 
\centering $\displaystyle\frac{x - 1}{\log x}$ \tabularnewline
\centering $\mathcal{E}(\lambda)$ & 
\centering $1/\lambda$ & 
\centering $-\lambda e^{\lambda} Ei(-\lambda)$\footnotemark & 
\centering $\displaystyle\frac{\lambda}{\lambda - \log x}$ \tabularnewline
\bottomrule[0.15 em]
\end{tabular}
\captionof{table}{\textit{Examples of different distributions.}}
\label{tab special cases}
\end{minipage}
\footnotetext[14]{$\mathcal{D}(\lambda) \equiv$ Dirac measure centred on $\lambda$, $\mathcal{U}([a,b]) \equiv$ uniform distribution over $[a,b]$, $\mathcal{E}(\lambda) \equiv$ exponential distribution with parameter $\lambda > 0$.}
\footnotetext[15]{For every $x < 0$, $Ei(x) := -\displaystyle\int_{-x}^{+\infty}{\frac{e^{-u}}{u}\,\mathrm{d}u}$.}

\begin{figure}[H]
\centering
\includegraphics[scale=0.40]{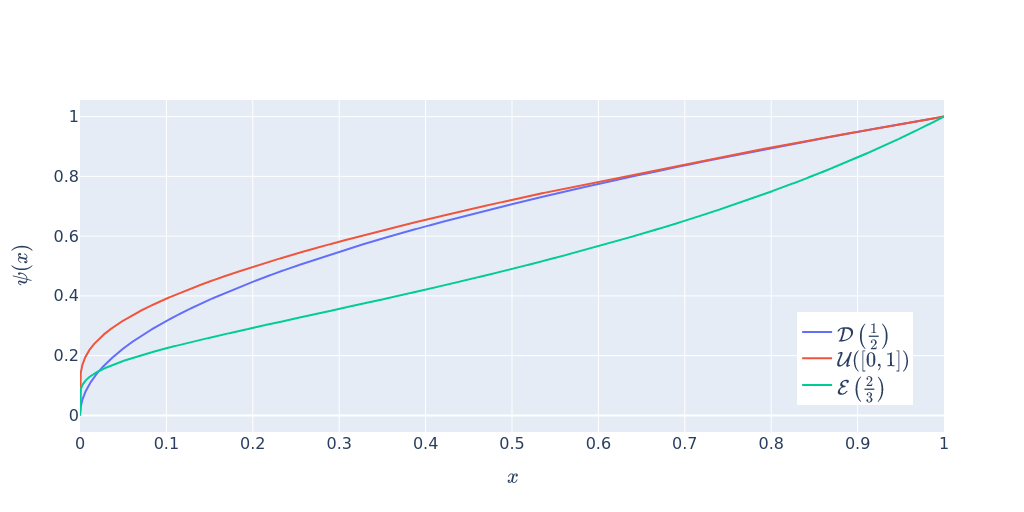}
\caption{Graph of the function $\psi$ for different distributions of $\rho$.}
\label{Figure special cases}
\end{figure}

\section{Excess Volatility and the Order-driven View of Markets}
\label{excess volatility and the order driven view of markets}

The volatility of stock prices is a well known phenomenon to all investors. Why, though, is this volatility so pronounced? As pointed by \cite{shiller1980stock} and \cite{leroy1981present}, it is apparent that there are extremely wide day-to-day changes in the prices quoted on most stock exchanges often exceeding the predictions of simple models with rational expectations. Shiller's argument is that the fluctuations are far too big to be accounted for by mere changes in information as confirmed in \cite{cutler1988moves} and \cite{fair2002events} for instance. In \cite{shiller1992market} the author introduces the (qualitative) popular model explanation of stock market volatility in which he proposes that investor reactions, due to psychological or sociological beliefs, exert a greater influence on the market than good economic sense arguments. In Shiller's eyes this excess volatility can be attributed to investors' psychological behaviour. He claims that substantial price changes can be explained by a collective change of mind by the investing public which can only be explained by its thoughts and beliefs on future events, i.e. its psychology\footnote{The mechanisms by which large returns have been obtained without significant changes in fundamentals are also well-described by the behavioral finance literature \cite{shleifer2000inefficient} \cite{hirshleifer2001investor} \cite{barberis2003survey}.}. The psychological effect of seeing others making a profit can be a powerful incentive that can cloud people's judgment and push them to follow each other. The story of how Isaac Newton allegedly lost £20,000 in the South Sea Bubble has become one of the most famous investment anecdote throughout history and illustrates this phenomenon. Back in the spring of 1720, Sir Isaac Newton owned shares in the South Sea Company, the hottest stock in England.  Newton dumped his South Sea shares, pocketing a profit totaling £20,000. But just months later, swept up in the wild enthusiasm of the market, Newton jumped back in at a much higher price and finally lost all the profits he made before \cite{odlyzko2020isaac}. In comparison, the average annual earning in the United Kingdom was around £13 at that time \cite{clark2022earnings}.

\begin{figure}[H]
\centering
\includegraphics[scale=0.9]{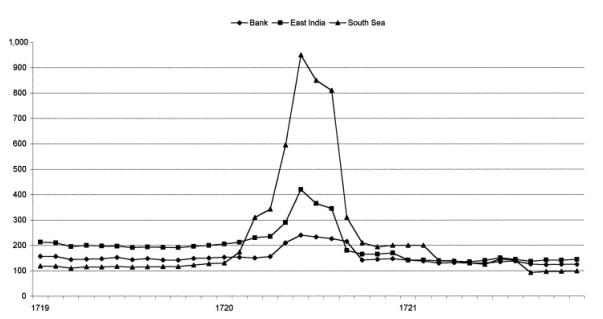}
\caption{The South Sea bubble. Three monthly share prices, 1719-21 from \cite{neal1991rise}.}
\label{Figure south sea bubble}
\end{figure}

We can a draw a parallel between these crowd effects and the apparent excess volatility that have been observed in the markets. Indeed, an aggregation of traders sharing the same belief can be viewed to some extent as a metaorder. Empirical evidences presented in Section \ref{metaorders market impact} are given on average. Hence Equation (\ref{average normalized market impact function}) and Table \ref{tab special cases} suggest that $\rho \sim \mathcal{D}\left(\frac{1}{2}\right)$ which gives that (\ref{corollary square root theorem equilibrium II}) can be written
\begin{equation}
\label{theoretical square-root law}
    \mathbb{E}\left[\varepsilon \left(S_{\tau_n} - S_{\tau_1^-}\right)\bigg|\mathcal{F}_n\right] \sim_{+\infty} \hat{\sigma}_n \sqrt{\frac{Q}{V}}.
\end{equation}
Furthermore, we have seen that empirically
\begin{equation}
\label{empirical square-root law}
    \mathbb{E}\left[\varepsilon \left(S_{\tau_n} - S_{\tau_1^-}\right)\bigg|\mathcal{F}_n\right] \sim_{+\infty} Y \sigma_n \sqrt{\frac{Q}{V}} 
\end{equation}
where $Y$ is a positive real random variable and $\sigma_n$ the volatility of the asset between $\tau_{1^-}$ and $\tau_n$. By combining (\ref{theoretical square-root law}) and (\ref{empirical square-root law}), we have
\begin{equation}
\label{order-driven relation}
    \hat{\sigma}_n \sim_{+\infty} Y \sigma_n.
\end{equation}
Denote the \textit{liquidity ratio} of the metaorder by $\mathcal{L}_n := \displaystyle\frac{\hat{\sigma}_n}{\sigma_n}$, $n \geq 1$, which measures the capacity of the market to absorb the metaorder, thus (\ref{order-driven relation}) states that the liquidity ratio converges $\mathbb{P}^{\mathscr{C}}-$a.s. towards $Y$, in line with the heuristic explanation\footnote{In Chapter 12, the authors define the liquidity ratio of the market over a time horizon $T$ (say in days) by $\mathcal{L}_T := \displaystyle\frac{\sqrt{V_T}}{\sigma_T}$ where $V_T$ and $\sigma_T$ stand for the traded volume and the volatility during $[0,T]$. They argue that if prices are exactly diffusive and the traded volume grows linearly with time, then $\mathcal{L}_T := \displaystyle\frac{\sqrt{V_T}}{\sigma_T} = \frac{\sqrt{T V_1}}{\sigma_1 \sqrt{T}} = \mathcal{L}_1$ and therefore the liquidity ratio is constant over a trajectory.} provided in \cite{bouchaud2018trades}. Recalling that $\hat{\sigma}_n$ is given by the market impact function over the trajectory of the metaorder taken in $V_1 + \dots + V_n$, i.e. the total volume traded between $\tau_{1^-}$ and $\tau_n$, (\ref{order-driven relation}) establishes a strong relation between the volume traded and the volatility. The conclusion is that, order flow, wheter informed or not, can be the major source of volatility in financial markets in line with \cite{gabaix2006institutional} and more recently the inelastic market story \cite{gabaix2021search}.

\section{Metaorders in Practice}
\label{metaorders in practice}

We introduce the discrete random variable $N$ taking only positive integers as values and representing the number of trades for a given metaorder, namely its \textit{length} to re-use the terminology introduced in \cite{said2017market}. Define $Q := \displaystyle\sum_{k=1}^{N}{Q_k}$ the \textit{size} of the metaorder.

\subsection{Metaorder Length and Size Distributions}
\label{metaorder length and size distributions}

In previous sections we have left the metaorder length and size distributions unspecified. Even if there is considerable accumulated evidence that in the large size limit in most major equity markets the metaorder length and size are distributed as power-laws (see Table \ref{tab length and size power law distributions}), we have seen that they are not involved in the establishment of the equilibrium. As a consequence, they do not play any role in the explanation of the shape of the impact. This is in line with recent empirical observations in the Bitcoin market \cite{donier2015million} and in contrast to some previous theories \cite{gabaix2003theory} \cite{farmer2013efficiency}.

\begin{center}
\begin{tabular}{l c c}
\toprule[0.15 em]
Empirical Study &
\multicolumn{1}{c}{$\xi_N$} & \multicolumn{1}{c}{$\xi_Q$} \\
\midrule[0.1 em]
\centering \cite{vaglica2008scaling} & 
\centering $1.80$ &
\centering $2.00$ \tabularnewline
\centering \cite{bershova2013non} & 
\centering $\varnothing$ &
\centering $1.48$ \tabularnewline
\centering \cite{said2017market} & 
\centering $1.40 - 1.80$ &
\centering $\varnothing$ \tabularnewline
\bottomrule[0.15 em]
\end{tabular}
\captionof{table}{\textit{$\xi_N$ and $\xi_Q$ such as $\mathbb{P}^{\mathscr{C}}(N > x) \sim_{+\infty} \displaystyle\frac{C_N}{x^{\xi_N}}$ and $\mathbb{P}^{\mathscr{C}}(Q > x) \sim_{+\infty} \displaystyle\frac{C_Q}{x^{\xi_Q}}$ with $C_N, C_Q > 0$. }}
\label{tab length and size power law distributions}
\end{center}

\begin{prop}[\textbf{$N$ and $Q$ distributions}]
\label{proposition asymptotic size distribution}
Let $\beta > 0$. If there exists $C > 0$ such as 
\begin{equation}
\label{asymptotic length distribution}
    \mathbb{P}^{\mathscr{C}}(N = n) \sim_{+\infty} \frac{C}{n^{1 + \beta}}
\end{equation}
then
$$ \mathbb{P}^{\mathscr{C}}\left(N \geq n+1\,|\, N \geq n\right) \sim_{+\infty} \left(1 + \frac{1}{n}\right)^{-\beta} $$
and there exist $C' > 0$, $C'' > 0$ and $M \geq 1$ such that for every $n \geq M$,
\begin{equation}
\label{asymptotic size distribution}
    \frac{C'}{n^{1 + \beta}} \leq \mathbb{P}^{\mathscr{C}}\left(n q_- \leq Q \leq n q_+\right) \leq \frac{C''}{n^{\beta}}.
\end{equation}
\end{prop}

\noindent Proposition \ref{proposition asymptotic size distribution} shows that the distribution of $Q$ is much more complicated to study. It is not even clear that $Q$ follows asymptotically a Pareto distribution when $N$ does and we have only (\ref{asymptotic size distribution}). This is mainly due to the fact that the metaorder size $Q$ strongly depends on its length $N$ and the lower and upper bounds of the size of its child order $q_-$ and $q_+$. 

\begin{prop}[\textbf{integrability of the effective pressure}]
\label{proposition convergence of the moments of the metaorder size}
Let $\beta > 0$, $\gamma \geq 0$ and $\nu \geq 0$. If (\ref{asymptotic length distribution}) holds and the conditional probability distribution of $Q$ given $N$ is a continuous uniform distribution over $\left[N q_-, N q_- + (q_+ - q_-)N^{1 - \gamma}\right]$ (resp. $\left[N q_+ - (q_+ - q_-)N^{1 - \gamma}, Nb\right]$), then $\mathbb{E}^{\mathscr{C}}\left[Q^{\nu}\right] < +\infty$ if and only if $\nu < \beta$.
\end{prop}

\noindent When the conditional probability distribution of $Q$ given $N$ is a continuous uniform distribution over $\left[N q_-, N q_- + (q_+ - q_-)N^{1 - \gamma}\right]$, the size of the child orders converges to $q_-$ when $N$ is large enough and the speed of this convergence is characterized by the parameter $\gamma$ since 
$$ N q_- + (q_+ - q_-)N^{1 - \gamma} = N\left[\left(1 - \displaystyle\frac{1}{N^{\gamma}}\right) q_- + \displaystyle\frac{1}{N^{\gamma}} q_+ \right]. $$
The same holds with $q_+$ in the case the conditional probability distribution of $Q$ given $N$ is a continuous uniform distribution over $\left[N q_+ - (q_+ - q_-)N^{1 - \gamma}, Nb\right]$. An interesting property underlined by Proposition \ref{proposition convergence of the moments of the metaorder size} is that the integrability of $Q^{\nu}$ is not affected by $\gamma$ but depends only on $\beta$. Considering that for a given metaorder $Q^{\rho}$ is the informational part of the impact we expect this quantity to be integrable. Write 
$$ \mathbb{E}^{\mathscr{C}}\left[Q^{\rho}\right] = \int_{0}^{+\infty}{\mathbb{E}^{\mathscr{C}}\left[Q^x\,|\,\rho = x\right]P_{\rho}(\mathrm{d}x)} $$
where for every Borel set A of $[0,+\infty)$, $P_{\rho}(A) := \mathbb{P}^{\mathscr{C}}(\rho \in A)$ and note that from Proposition \ref{proposition convergence of the moments of the metaorder size} if $\mathbb{E}^{\mathscr{C}}\left[Q^{\rho}\right] < +\infty$, then $\mathbb{P}^{\mathscr{C}}\left(\rho < \beta\right) = 1$. This gives a strong interpretation to $\beta$ as an uniform upper bound of the random variable $\rho$ under assumptions of Proposition \ref{proposition convergence of the moments of the metaorder size}.

\subsection{Concave and Non-decreasing Market Impact}
\label{concave and non-decreasing market impact}

\noindent For large investors, it is not enough to know the current price of an asset to determine the best plan for buying or selling — the likely effect of the order itself must be taken into consideration. Although this concept may seem obvious and is certainly evident in actual trading data, the fact that market impact can increase trading costs is an important factor in how large trades are accomplished — particularly for crowded strategies. Since the 1990s, monitoring and controlling market impact has become an active area of research in quantitative finance, encompassing the size and timing of trades, cross-impacts and market liquidity. With that in mind, one of the central questions in constructing any large trade will be the slippage impact of that trading — how much will it cost and is the market liquid enough to accommodate it immediately? For many substantial orders, it is necessary to \textit{slice and dice} the order — arranging for execution over several days rather than all at once. However, during this time, the price may move — in part due to natural fluctuations in the market and in part due to the impact of the trade itself. Whether the price moves a little or a lot depends on the quantity and the timing. For sophisticated traders working with good signals, it may be possible to predict and mitigate the move to some extent, but, clearly, execution risk can be painful if the price moves sharply against the trader. During the 1980s, a number of authors and market participants began producing research on this phenomenon e.g \cite{loeb1983trading}. An empirical investigation of the market impact has shown that such impact appears to be governed by an increasing concave functional under reasonable trading regime and market conditions, with nonlinear impact and decay over time.

\begin{prop}[\textbf{upper and lower bounds for $\mathcal{R}_N$}]
\label{real metaorders and non-decreasing concave market impact functions}
Assume that the conditions of Theorem \ref{theorem equilibrium II} hold with $\rho > 0$ and set for every $x \geq 0$, $f(x) := x^{\rho} \exp\left(\eta(x) + \displaystyle\int_{0}^{x}{\frac{\theta(u)}{u}\mathrm{d}u}\right)$. If $f$ is non-decreasing and concave, then $\mathbb{P}^{\mathscr{C}}-$a.s. 
$$ \frac{1}{2} \leq \mathcal{R}_N \leq 1. $$
\end{prop}

\noindent Proposition \ref{real metaorders and non-decreasing concave market impact functions} makes some predictions about the values taken by $\mathcal{R}_N$ when the market impact function is concave and non-decreasing. Having in mind that $\mathcal{R}_N \in [0,1]$ can be seen as a performance measure of the execution of the metaorder (see. Section \ref{rho = infinity}), this is in line with the empirical observations: Under reasonable market conditions (which corresponds to our event $\mathscr{C}$) market impact is a concave and increasing phenomenon and the market participants accept to trade only if the quotes offered by the market makers are in their favor. We will see in the next section how these upper and lower bounds can be taken into account to shed some light on the reversion process of the metaorder.  

\subsection{Metaorder Relaxation and Fair Pricing Point}
\label{metaorder relaxation and fair pricing point}

Giving a metaorder we introduce its \textit{fair pricing point}\footnote{It is also the point chosen in \cite{farmer2013efficiency} to set the permanent market impact in their model.} as the point when the price trajectory goes back to its average execution price just after it has been fully executed. We introduce also its \textit{fair pricing time} as the time needed for the price trajectory to reach the fair pricing point. Set $(\mathcal{N}_t)_{t \geq 0}$ a noise term such that for all $t \geq 0$, $\mathbb{E}^{\mathscr{C}}\left[\mathcal{N}_t\right] = 0$ and define the \textit{relaxation function} $\mathcal{G}_N$ up to $t$ and the \textit{average relaxation function} $G$ on $[0,+\infty)$ by
$$ \mathcal{G}_N(t) := \frac{\mathbb{E}\left[\varepsilon \left(S_{\tau_N + t} - S_{\tau_1^-}\right)\bigg|\mathcal{F}_N\right]}{\mathbb{E}\left[\varepsilon \left(S_{\tau_N} - S_{\tau_1^-}\right)\bigg|\mathcal{F}_N\right]} + \mathcal{N}_t $$
and
\begin{equation}
\label{average relaxation function}
    G(t) := \mathbb{E}^{\mathscr{C}}\left[\mathcal{G}_N(t)\right].
\end{equation}

\noindent Note that $t=0$ corresponds to the end of the execution of the metaorder and the start of its relaxation. The shape of the average relaxation function $G$ has been empirically investigated in \cite{moro2009market} \cite{bershova2013non} \cite{brokmann2015slow} \cite{bacry2015market} \cite{said2017market} and \cite{bucci2018slow} in the equity market and \cite{said2021market} in the options market. All these studies have found that the relaxation function is convex decreasing.

\begin{prop}[\textbf{upper bounds for the fair pricing time}]
\label{proposition metaorder duration inequalities}
Assume that the conditions of Proposition \ref{real metaorders and non-decreasing concave market impact functions} hold, consider $G$ positive, continuous and decreasing such that $\displaystyle\frac{1}{2} \geq \alpha$ where $ \alpha := G(\infty) = \displaystyle\inf_{t \geq 0}{G(t)} \geq 0$ and define $T_N := G^{-1}(\mathcal{R}_N)$ the fair pricing time. Then we have
\begin{equation}
\label{metaorder duration inequality}
    0 \leq T_N \leq G^{-1}\left(\frac{1}{2}\right).
\end{equation}
Furthermore, if $G$ is convex
\begin{equation}
\label{average metaorder duration inequality}
    \mathbb{E}^{\mathscr{C}}\left[T_N\right] \leq G^{-1}\left(\mathbb{E}^{\mathscr{C}}\left[\mathcal{R}_N\right]\right).
\end{equation}
\end{prop}

\noindent Under assumptions of Proposition \ref{proposition metaorder duration inequalities} $\alpha$ can be seen as the long time information content of the trades. Define the \textit{uninformed average relaxation function} $G_0$ on $[0,+\infty)$ by
$$ G_0(t) := \frac{G(\infty) - G(t)}{G(\infty) - G(0)}, $$
hence $G$ may be rewritten
$$ G(t) = \alpha + (1 - \alpha)G_0(t). $$
Note that $G_0$ and $G$ have the same properties except that $\lim\limits_{t \rightarrow +\infty}{G_0(t)} = 0$ whereas $\lim\limits_{t \rightarrow +\infty}{G(t)} = \alpha$. Hence, once the informational part of the trade is removed, what remains converges to $0$ in agreement with \cite{brokmann2015slow} \cite{bacry2015market}. Furthermore, $\lim\limits_{\eta \rightarrow \alpha^+}{G^{-1}(\eta)} = +\infty$ indicates that the decay may be slow \cite{brokmann2015slow} \cite{bucci2018slow}. For a given metaorder in the equilibrium case $T_N$ is the time needed for the price trajectory to reach the fair pricing point. Proposition \ref{proposition metaorder duration inequalities} shows that when the market impact function is concave, $T_N$ and $\mathbb{E}^{\mathscr{C}}[T_N]$ admit uniform upper bounds given by (\ref{metaorder duration inequality}) and (\ref{average metaorder duration inequality}). Define the \textit{residual market impact function} on $[0,+\infty)$ by
\begin{equation}
\label{residual market impact function}
    I_N(t) := I(Q_1 + \dots + Q_N, t) = G(t)\mathcal{I}_N.
\end{equation}
Note that
\begin{equation}
\label{residual market impact at the fair pricing point}
    I_N(T_N) = G(T_N)\mathcal{I}_N = \langle \mathcal{I} \rangle_N
\end{equation}
and
\begin{equation}
\label{residual market impact at infinity}
    I_N(\infty) := \lim\limits_{t \rightarrow +\infty}{I_N(t)} = \alpha \mathcal{I}_N.
\end{equation}
In what follows we will say that 
\begin{itemize}
    \item (\ref{residual market impact at the fair pricing point}) is the \textit{residual market impact at the fair pricing point},
    \item (\ref{residual market impact at infinity}) is the \textit{residual market impact at} $\infty$.
\end{itemize}
Two interesting points are 
\begin{itemize}
    \item the residual market impact at the fair pricing point (\ref{residual market impact at the fair pricing point}) and at $\infty$ (\ref{residual market impact at infinity}) follow roughly the same functional form. For instance, if (\ref{residual market impact at the fair pricing point}) follows a square-root law as verified by \cite{bershova2013non}, then the same holds for (\ref{residual market impact at infinity});
    \item the residual market impact at $\infty$ is proportional to the temporary impact $\mathcal{I}_N$.
\end{itemize}

\subsection{Different Stories of Permanent Market Impact}
\label{different stories of permanent market impact}

\noindent If it is widely recognized that temporary impact has a concave dependence on size \cite{almgren2005direct} \cite{engle2012measuring} \cite{bacry2015market}, the functional form of permanent impact is harder to measure and remains an open question. Especially differences arise in the price reversion following the end of a metaorder. The existing empirical literature of decay metaorders market impact is limited \cite{moro2009market} \cite{bershova2013non} \cite{brokmann2015slow}  \cite{gomes2015market} \cite{bacry2015market} \cite{said2017market} \cite{bucci2018slow} due to the difficulty of obtaining data. This second type of market impact is more controversial and research papers dealing with permanent market impact can be separated in two categories. On one hand the permanent market impact can be seen as the consequence of a pure mechanical process. On the other hand the permanent market impact is considered to be a trace of new information in the price. In the pure mechanical vision prices move because of the activity of all the market participants. So if the buy pressure takes advantage on the sell pressure the price go up, and if selling pressure is stronger than buying pressure the price go down. This is the econophysicist point of view which main goal is to determine the behavior of these two forces and how they generate impact on prices' dynamic. The second school of thought is the economist point of view: The informational vision says prices move because new information is made available to investors who update their expectations. As a consequence the market participants change their offer and demand which gives birth to a global new equilibrium resulting in new prices levels. In this picture, as emphasized in \cite{hasbrouck2007empirical}, orders do not impact prices and it is more accurate to say that orders forecast prices. Among those whose share the mechanical vision of the permanent impact there are also two pictures. On one side, there is the framework proposed by \cite{bouchaud2010price} where there is no such thing as permanent impact but only the long memory of the sign of the metaorder flow. On the other side, the picture of \cite{farmer2013efficiency} states that permanent impact can be important and roughly equals to $2/3$ of the peak impact. This is the fair pricing hypothesis. A range of papers have analyzed all sorts of metaorder databases reaching conclusions in favour of one position or the other as summarized in Table \ref{tab decay empirical studies}.

\begin{center}
\begin{tabular}{l l}
\toprule[0.15 em]
Empirical study & \multicolumn{1}{c}{Residual impact / Temporary
impact} \\
\midrule[0.1 em]
\centering \cite{moro2009market} & \centering $0.5 \sim 0.7$ (single day metaorders) \tabularnewline
\centering \cite{bershova2013non} & \centering $\sim 2/3$ (single day metaorders) \tabularnewline
\centering \cite{gomes2015market} & \centering $\sim 2/3$ (informed) -- $\sim 0$ (uninformed) after $10$ days \tabularnewline
\centering \cite{said2017market} & \centering $\sim 2/3$ (single day metaorders) \tabularnewline
\centering \cite{bucci2018slow} & \centering $\sim 2/3$ at the end of the same day -- $\sim 1/3$ after $50$ days \tabularnewline
\bottomrule[0.15 em]
\end{tabular}
\captionof{table}{\textit{Decay of the impact obtained in some empirical studies}}
\label{tab decay empirical studies}
\end{center}

As far as empirical data is concerned, the situation is also rather confusing, mostly because the determination of the time when the relaxation is studied varies from one study to another. Even the formal definition of what is called permanent market impact is not clear. Some authors take as definition of permanent impact (\ref{residual market impact at the fair pricing point}) \cite{farmer2013efficiency} whereas others prefer consider (\ref{residual market impact at infinity}) \cite{donier2015fully}. Furthermore, the terminology \textit{permanent market impact} is misleading by inducing the idea that permanent impact is as universal as its temporary counterpart. We will prefer to use the term of residual market impact after $t \in [0,+\infty)$ (units of time) to refer to the decay of the impact after the metaorder has been filled. The traditional view in finance is that market impact is just a reflection of information and postulates that the functional form of market impact is the expression of how informed the agents are who trade with a given volume. As information is difficult to define and measure the metaorder size has been used as an explanatory variable for the temporary market impact. If it seems reasonable to assume that the characteristics of metaorders can determine the shape of the temporary impact, they cannot explain what happens next: Once executed, the information reflected in the metaorder is subject to market noise. So the residual market impact must be the result of this interference and justifies that relaxation can be only be taken on consideration on average as expressed in (\ref{average relaxation function}). Note also that the residual market impact at $\infty$ depends on $\alpha$ which is strongly correlated by definition to the set of metaorders used to estimate $G$ in agreement with the empirical observations in \cite{gomes2015market}.

The value 2/3 appears persistently in several studies (see Table \ref{tab decay empirical studies}) and recently has been a subject of controversy. In \cite{bucci2018slow}, the authors have analyzed more than $8$ millions metaorders executed by institutional investors in the US equity market and shown that even if the relaxation the same day is on average $\approx 2/3$ of the peak impact, the decay continues the next days. This value corresponds to the fair pricing point of our model. Set $X_N\footnote{We adopt here the same notation as in \cite{farmer2013efficiency}.} := S_{\tau_N + T_N}$ the price at the fair pricing point i.e.
$$ X_N = \frac{\displaystyle \sum_{k=1}^{N}{Q_k\, \mathbb{E}[S_{\tau_k}\,|\,\mathcal{F}_k]}}{\displaystyle \sum_{k=1}^{N}{Q_k}}. $$
In \cite{farmer2013efficiency} the permanent market impact is defined as
$$ I_N^{\text{Farmer}} := \varepsilon(X_N - S_{\tau_1^-}) = \langle \mathcal{I} \rangle_N = I_N(T_N), $$
thus the permanent market impact in the Farmer's model corresponds to the residual market impact at the fair pricing point as defined in (\ref{residual market impact at the fair pricing point}). And indeed, at the fair pricing point $I_N(T_N) = \mathcal{R}_N \mathcal{I}_N \approx \displaystyle\frac{2}{3} \mathcal{I}_N$ when $N$ is large enough and $\rho = \displaystyle\frac{1}{2}$. This is another evidence in favor of the square-root behavior of market impact. More precisely, \cite{said2017market} and \cite{said2021market} have suggested that empirically $\mathbb{E}^{\mathscr{C}}\left[T_N\right] \approx \mathbb{E}^{\mathscr{C}}\left[\tau_N - \tau_{1^-}\right]$ which means that on average the time needed to reach the fair pricing point is almost equal to the duration of the metaorder.

\section{Conclusion}
\label{conclusion}

This paper presents a theory of the market impact of metaorders based on a supply and demand equilibrium replicating most of the stylized facts that have been observed in the empirical literature. An equilibrium is reached when the ratio between the average impact and the peak impact converges to $\displaystyle\frac{1}{1 + \rho}$ shaping the temporary market impact as $Q^{\rho} \ell(Q)$ where $\ell$ is a slowly varying function and $Q$ the metaorder size. Our model reproduces the square-root law, predicts non-trivial price trajectories in non-equilibrium situations and sheds some light on the excess volatility puzzle. Furthermore, we have shown that according to empirical evidences the random variable $\rho$ likely follows a Dirac measure centred on $1/2$, which means that the price trajectory of a given metaorder converges also to a square-root law when the metaorder size is large enough. This also reinforces the idea that the market impact of metaorders is ruled by universal mechanisms.

\section*{Acknowledgements}

We thank Marcos Lopez de Prado and Alexander Lipton for their comments on the preliminary version of this paper. The author is particularly grateful to Charles-Albert Lehalle for his careful reading, comments and the many interesting discussions we shared.

\section{Proofs}

\begin{lem}
\label{lemma 1}
Let $(\alpha_n)_{n \geq 1}$ a sequence of positive real numbers. If 
$$ \lim\limits_{n \rightarrow +\infty}{\frac{Q_1 + \dots + Q_n}{Q_n}\left(1 - \frac{\alpha_{n-1}}{\alpha_n}\right)} = \rho \in [0,+\infty], $$ 
then 
$$ \lim\limits_{n \rightarrow +\infty}{\frac{\displaystyle\sum_{k=1}^{n}{Q_k \alpha_k}}{\left(\displaystyle\sum_{k=1}^{n}{Q_k}\right) \alpha_n}} = \frac{1}{1 + \rho}. $$
\end{lem}

\begin{lem}
\label{lemma 2}
Let $(\alpha_n)_{n \geq 1}$ such that for all $n \geq 1$,
    $$ \alpha_n = (Q_1 + \dots + Q_n)^{\rho}\exp\left(\eta + \int_{0}^{Q_1 + \dots + Q_n}{\frac{\epsilon(u)}{u}\,\mathrm{d}u}\right) $$
    where $\rho \geq 0$, $\eta \in \mathbb{R}$ and $\theta$ is a bounded measurable function of a real variable with $\lim\limits_{x \rightarrow +\infty}{\theta(x)} = 0$. Then we have
    $$ \frac{\alpha_{n-1}}{\alpha_n} = 1 - \rho\frac{Q_n}{Q_1 + \dots + Q_n} + o\left(\frac{Q_n}{Q_1 + \dots + Q_n}\right) $$
    as $n \rightarrow +\infty$.
\end{lem}

\subsection{Proof of Lemma \ref{lemma 1}}

We proceed in four steps.

\begin{itemize}
    \item We prove first that for all $\rho \in [0,+\infty)$, $\limsup\limits_{n \rightarrow +\infty}{\mathcal{Z}_n} \leq \displaystyle\frac{1}{1 + \rho}$.
    \item Then we prove that for all $\rho \in [0,+\infty)$, $\liminf\limits_{n \rightarrow +\infty}{\mathcal{Z}_n} \geq \displaystyle\frac{1}{1 + \rho}$.
    \item We conclude for every $\rho \in [0,+\infty)$.
    \item We extend the result to $\rho = +\infty$.
\end{itemize}

\begin{proof}
Set for all $n \in \mathbb{N}^*$, 
$$ \mathcal{Z}_n := \frac{\displaystyle\sum_{k=1}^{n}{Q_k \alpha_k}}{\left(\displaystyle\sum_{k=1}^{n}{Q_k}\right) \alpha_n}, $$
and for every $n \geq 2$, $\Delta_n := \displaystyle\frac{\alpha_{n-1}}{\alpha_n}$, $\varepsilon_n := 1 - \Delta_n$.
\begin{itemize}
    \item Assumption \ref{assumption a.s. VWAP constraint} implies that the result holds when $\rho = 0$. Let $\rho \in (0, +\infty)$ such that $\lim\limits_{n \rightarrow +\infty}{\displaystyle\frac{Q_1 + \dots + Q_n}{Q_n} \varepsilon_n} = \rho$. Let $\varepsilon \in (0,\rho)$. Set for any $n \geq 2$, $W_n := \displaystyle\frac{Q_1 + \dots + Q_n}{Q_n} \varepsilon_n$. There exists $N \geq 2$ such that for all $n > N$, $\left|W_n - \rho\right| \leq \varepsilon$. For every $n \geq 2$,
    $$ \mathcal{Z}_n \left(\sum_{k=1}^{n}{Q_k}\right) = \mathcal{Z}_{n-1} \left(\sum_{k=1}^{n-1}{Q_k}\right)\Delta_n + Q_n, $$
    and then by a straightforward induction for all $n > N$,
    \begin{align*}
        \mathcal{Z}_n \left(\sum_{k=1}^{n}{Q_k}\right) &= \mathcal{Z}_{N-1}\sum_{k=1}^{N-1}{Q_k \prod_{i=k+1}^{n}{\Delta_i}} + \sum_{k=N}^{n}{Q_k \prod_{i=k+1}^{n}{\Delta_i}} \\
        &\leq C \mathcal{Z}_{N-1} \sum_{k=1}^{N-1}{Q_k} + \sum_{k=N}^{n}{Q_k \prod_{i=k+1}^{n}{(1 - \varepsilon_i})}
    \end{align*}
    where $C := \displaystyle\prod_{n\,:\,\Delta_n \geq 1}{\Delta_n} < +\infty$ since $\{n \geq 2\,|\,\Delta_n \geq 1\}$ is a finite set and with the convention that a product with no terms is equal to 1. Similarly, in what follows, a sum with no terms will evaluate to $0$. Let $n > N$.
    \begin{align*}
        \sum_{k=N}^{n}{Q_k \prod_{i=k+1}^{n}{(1 - \varepsilon_i})} &= \sum_{k=N}^{n}{Q_k \exp\left(\sum_{i=k+1}^{n}{\log(1 - \varepsilon_i)}\right)} \\
        &= \sum_{k=N}^{n}{Q_k \exp\left(\sum_{i=k+1}^{n}{\log\left(1 - \frac{Q_i}{Q_1 + \dots + Q_i} W_i\right)}\right)} \\
        &\leq \sum_{k=N}^{n}{Q_k \exp\left(\sum_{i=k+1}^{n}{\log\left(1 - \frac{Q_i}{Q_1 + \dots + Q_i} (\rho - \varepsilon)\right)}\right)} \\
        &\leq \sum_{k=N}^{n}{Q_k \exp\left(-(\rho - \varepsilon)\sum_{i=k+1}^{n}{\frac{Q_i}{Q_1 + \dots + Q_i}}\right)} \\
        &\leq \sum_{k=N}^{n}{Q_k\exp\left(-(\rho - \varepsilon)\int_{Q_1 + \dots + Q_k}^{Q_1 + \dots + Q_{n+1}}{\frac{\mathrm{d}x}{x}}\right)} \\
        &\leq \sum_{k=N}^{n}{Q_k\exp\left(-(\rho - \varepsilon)\log\left(\frac{\mathcal{S}_{n+1}}{\mathcal{S}_k}\right)\right)} \\
        &\leq \sum_{k=N}^{n}{Q_k\left(\frac{\mathcal{S}_k}{\mathcal{S}_{n+1}}\right)^{\rho - \varepsilon}}
    \end{align*}
    by setting $\mathcal{S}_n := \displaystyle\sum_{k=1}^{n}{Q_k}$ for every $n \in \mathbb{N}$. Hence for all $n > N$,
    \begin{align*}
        \mathcal{Z}_n &\leq C\mathcal{Z}_{N-1}\frac{\displaystyle\sum_{k=1}^{N-1}{Q_k}}{\displaystyle\sum_{k=1}^{n}{Q_k}} + \sum_{k=N}^{n}{\left(\frac{\mathcal{S}_k}{\mathcal{S}_n} - \frac{\mathcal{S}_{k-1}}{\mathcal{S}_n}\right)}\left(\frac{\mathcal{S}_k}{\mathcal{S}_{n+1}}\right)^{\rho - \varepsilon} \\
        &\leq C\mathcal{Z}_{N-1}\frac{\displaystyle\sum_{k=1}^{N-1}{Q_k}}{\displaystyle\sum_{k=1}^{n}{Q_k}} + \sum_{k=1}^{n}{\left(\frac{\mathcal{S}_k}{\mathcal{S}_n} - \frac{\mathcal{S}_{k-1}}{\mathcal{S}_n}\right)\left(\frac{\mathcal{S}_k}{\mathcal{S}_n}\right)^{\rho - \varepsilon}}
    \end{align*}
    whence for all $\varepsilon > 0$, $\limsup\limits_{n \rightarrow +\infty}{\mathcal{R}_n} \leq \displaystyle\frac{1}{1 + \rho - \varepsilon}$ and
    \begin{equation}
    \label{limsup inequality}
        \limsup\limits_{n \rightarrow +\infty}{\mathcal{Z}_n} \leq \displaystyle\frac{1}{1 + \rho}.
    \end{equation}
    \item Let $\varepsilon > 0$. There exists $N \geq 2$ such that
        \begin{itemize}
            \item $\forall \,n > N$, $\left|W_n - \rho\right| \leq \varepsilon$,
            \item $\forall \,p,q > N$, $\displaystyle\left|\left(\frac{Q_p}{Q_1 + \dots + Q_p}\right)^2 + \cdots + \left(\frac{Q_q}{Q_1 + \dots + Q_q}\right)^2\right| \leq \varepsilon$,
            \item $\forall \,n > N$, $\displaystyle\frac{Q_n}{Q_1 + \dots + Q_n}(\rho + \varepsilon) < \displaystyle\frac{1}{2}$.
        \end{itemize}
        For every $n > N$,
        \begin{align*}
            \mathcal{Z}_n \left(\sum_{k=1}^{n}{Q_k}\right) &= \mathcal{Z}_{N-1}\sum_{k=1}^{N-1}{Q_k \prod_{i=k+1}^{n}{\Delta_i}} + \sum_{k=N}^{n}{Q_k \prod_{i=k+1}^{n}{\Delta_i}} \\
            &\geq \sum_{k=N}^{n}{Q_k \prod_{i=k+1}^{n}{(1 - \varepsilon_i})}
        \end{align*}
        \begin{align*}
            \sum_{k=N}^{n}{Q_k \prod_{i=k+1}^{n}{(1 - \varepsilon_i})} &= \sum_{k=N}^{n}{Q_k \exp\left(\sum_{i=k+1}^{n}{\log(1 - \varepsilon_i)}\right)} \\
            &= \sum_{k=N}^{n}{Q_k \exp\left(\sum_{i=k+1}^{n}{\log\left(1 - \frac{Q_i}{Q_1 + \dots + Q_i} W_i\right)}\right)} \\
            &\geq \sum_{k=N}^{n}{Q_k \exp\left(\sum_{i=k+1}^{n}{\log\left(1 - \frac{Q_i}{Q_1 + \dots + Q_i} (\rho + \varepsilon)\right)}\right)} \\
            &\geq \sum_{k=N}^{n}{Q_k \exp\left(- (\rho + \varepsilon)\sum_{i=k+1}^{n}{\frac{Q_i}{Q_1 + \dots + Q_i}} - (\rho + \varepsilon)^2 \sum_{i=k+1}^{n}{\left(\frac{Q_i}{Q_1 + \dots + Q_i}\right)^2}\right)} \\
            &\geq e^{-(\rho + \varepsilon)^2 \varepsilon} \sum_{k=N}^{n}{Q_k \exp\left(-(\rho + \varepsilon)\sum_{i=k+1}^{n}{\frac{Q_i}{Q_1 + \dots + Q_i}}\right)} \\
            &\geq e^{-(\rho + \varepsilon)^2 \varepsilon} \sum_{k=N}^{n}{Q_k\exp\left(-(\rho + \varepsilon)\int_{Q_1 + \dots + Q_k}^{Q_1 + \dots + Q_n}{\frac{\mathrm{d}x}{x}}\right)} \\
            &\geq e^{-(\rho + \varepsilon)^2 \varepsilon} \sum_{k=N}^{n}{Q_k\exp\left(-(\rho + \varepsilon)\log\left(\frac{\mathcal{S}_n}{\mathcal{S}_k}\right)\right)} \\
            &\geq  e^{-(\rho + \varepsilon)^2 \varepsilon} \sum_{k=N}^{n}{Q_k\left(\frac{\mathcal{S}_k}{\mathcal{S}_n}\right)^{\rho + \varepsilon}}
        \end{align*}
    where we have used that for all $x \in [0,1/2)$, $\ln(1-x) \geq -x - x^2$. Thus for each $n > N$,
    $$ \mathcal{Z}_n \geq e^{-(\rho + \varepsilon)^2 \varepsilon} \sum_{k=1}^{n}{\left(\frac{\mathcal{S}_k}{\mathcal{S}_n} - \frac{\mathcal{S}_{k-1}}{\mathcal{S}_n}\right)\left(\frac{\mathcal{S}_k}{\mathcal{S}_n}\right)^{\rho + \varepsilon}} - e^{-(\rho + \varepsilon)^2 \varepsilon} \frac{\displaystyle\sum_{k=1}^{N-1}{Q_k\left(\frac{\mathcal{S}_k}{\mathcal{S}_n}\right)^{\rho + \varepsilon}}}{\displaystyle\sum_{k=1}^{n}{Q_k}}. $$
    Hence for all $\varepsilon > 0$, $\liminf\limits_{n \rightarrow +\infty}{\mathcal{Z}_n} \geq \displaystyle\frac{e^{-(\rho + \varepsilon)^2 \varepsilon}}{1 + \rho +\varepsilon}$ which gives that
    \begin{equation}
    \label{liminf inequality}
        \liminf\limits_{n \rightarrow +\infty}{\mathcal{Z}_n} \geq \displaystyle\frac{1}{1 + \rho}.
    \end{equation}
\item From (\ref{limsup inequality}) and (\ref{liminf inequality}) we have
$$ \limsup\limits_{n \rightarrow +\infty}{\mathcal{Z}_n} = \liminf\limits_{n \rightarrow +\infty}{\mathcal{Z}_n}= \displaystyle\frac{1}{1 + \rho}, $$
which gives the convergence of the sequence $(\mathcal{Z}_n)_{n \geq 1}$ to the limit $\displaystyle\frac{1}{1 + \rho}$ for each $\rho \in [0,+\infty)$.
\item Suppose $\rho = +\infty$. Let $\rho' \geq 0$, $\left(\varepsilon'_n\right)_{n \geq 2}$ and $\left(\mathcal{Z}'_n\right)_{n \geq 1}$ two sequences defined such that for every $n \geq 1$,
$$ \varepsilon'_n = \min\left(\varepsilon_n, \frac{Q_n}{Q_1 + \dots + Q_n}\rho'\right) $$
and
$$
\begin{cases}
    \mathcal{Z}'_1 = 1, \\
    \forall n \geq 1, \mathcal{Z}'_{n+1} = \displaystyle\frac{Q_{n+1} + (Q_1 + \dots + Q_n) \mathcal{Z}'_n (1-\varepsilon'_{n+1})}{Q_1 + \dots + Q_{n+1}}
\end{cases}
$$
Recalling that $(\mathcal{Z}_n)_{n \geq 1}$ is given by
$$
\begin{cases}
  \mathcal{Z}_1 = 1, \\
  \forall n \in \geq 1, \mathcal{Z}_{n+1} =  \displaystyle\frac{Q_{n+1} + (Q_1 + \dots + Q_n) \mathcal{Z}_n (1-\varepsilon_{n+1})}{Q_1 + \dots + Q_{n+1}},
\end{cases}
$$
we get that for all $n \geq 1$, $\mathcal{Z}'_n \geq \mathcal{Z}_n$, leading to $\limsup\limits_{n \rightarrow +\infty}{\mathcal{Z}'_n} \geq \limsup\limits_{n \rightarrow +\infty}{\mathcal{Z}_n}$. Furthermore since $\lim\limits_{n \rightarrow +\infty}{\displaystyle\frac{Q_1 + \dots + Q_n}{Q_n} \varepsilon'_n} = \rho'$, we have $\lim\limits_{n \rightarrow +\infty}{\mathcal{Z}'_n} = \displaystyle\frac{1}{1 + \rho'}$. Hence for all $\rho' \geq 0$, $\displaystyle\frac{1}{1 + \rho'} \geq \limsup\limits_{n \rightarrow +\infty}{\mathcal{Z}_n}$. This leads to $\limsup\limits_{n \rightarrow +\infty}{\mathcal{Z}_n} = 0$.
\end{itemize}
\end{proof}

\subsection{Proof of Lemma \ref{lemma 2}}

\begin{proof}
For all $n \geq 2$,
\begin{equation}
\label{eqn 1 proof lemma 2}
    \frac{\alpha_{n-1}}{\alpha_n} = \left(1 - \frac{Q_n}{Q_1 + \dots + Q_n}\right)^{-\rho}\exp\left(-\int_{Q_1 + \dots + Q_{n-1}}^{Q_1 + \dots + Q_n}{\frac{\theta(u)}{u}\,\mathrm{d}u}\right).
\end{equation}
Let $\varepsilon > 0$. There exist $A \geq 0$, $N \in \mathbb{N}^*$ such that for every $x \geq A$, $n > N$, $|\theta(x)| \leq \varepsilon$ and $Q_1 + \dots + Q_{n-1} \geq A$, hence for every $n > N$,
$$ \left|\int_{Q_1 + \dots + Q_{n-1}}^{Q_1 + \dots + Q_n}{\frac{\theta(u)}{u}\,\mathrm{d}u}\right| \leq \int_{Q_1 + \dots + Q_{n-1}}^{Q_1 + \dots + Q_n}{\frac{\varepsilon}{u}\,\mathrm{d}u} \leq \varepsilon \frac{Q_n}{Q_1 + \dots + Q_{n-1}}$$
which gives
$$ \frac{Q_1 + \dots + Q_n}{Q_n} \left|\int_{Q_1 + \dots + Q_{n-1}}^{Q_1 + \dots + Q_n}{\frac{\theta(u)}{u}\,\mathrm{d}u}\right| \leq \varepsilon \left(1 - \frac{Q_n}{Q_1 + \dots + Q_n}\right)^{-1}. $$
Thus
$$ \int_{Q_1 + \dots + Q_{n-1}}^{Q_1 + \dots + Q_n}{\frac{\theta(u)}{u}\,\mathrm{d}u} = o\left(\frac{Q_n}{Q_1 + \dots + Q_n}\right) \text{as} \enspace n \rightarrow +\infty$$
and 
\begin{align*}
    \exp\left(-\int_{Q_1 + \dots + Q_{n-1}}^{Q_1 + \dots + Q_n}{\frac{\theta(u)}{u}\,\mathrm{d}u}\right) &= 1 - \int_{Q_1 + \dots + Q_{n-1}}^{Q_1 + \dots + Q_n}{\frac{\theta(u)}{u}\,\mathrm{d}u} + o\left(\int_{Q_1 + \dots + Q_{n-1}}^{Q_1 + \dots + Q_n}{\frac{\theta(u)}{u}\,\mathrm{d}u}\right) \\
    &= 1 + o\left(\frac{Q_n}{Q_1 + \dots + Q_n}\right) \numberthis \label{eqn 2 proof lemma 2}
\end{align*}
as $n \rightarrow +\infty$. Then by plugging (\ref{eqn 2 proof lemma 2}) in (\ref{eqn 1 proof lemma 2}), finally we have
$$ \frac{\alpha_{n-1}}{\alpha_n} = 1 - \rho\frac{Q_n}{Q_1 + \dots + Q_n} + o\left(\frac{Q_n}{Q_1 + \dots + Q_n}\right) $$
as $n \rightarrow +\infty$.
\end{proof}

\subsection{Proof of Theorem \ref{theorem equilibrium I}}

We proceed in three steps by proving $(iii) \iff (iv)$, $(i) \iff (iii)$ and $(ii) \iff (iv)$.

\begin{proof}
Set $\alpha_n := \left[\varepsilon\left(S_{\tau_n} - S_{\tau_1^-}\right)\bigg|\mathcal{F}_n\right]$, $n \geq 1$.
\begin{itemize}
    \item $(iii) \iff (iv)$  Let $\sigma < \rho$ and $\tau > \rho$.
    \begin{itemize}
        \item $(iii) \implies (iv)$ We have
        \begin{align*}
            &\quad\,(Q_1 + \dots + Q_n)^{-\sigma} \alpha_n - (Q_1 + \dots + Q_{n-1})^{-\sigma} \alpha_{n-1} \\
            &= (Q_1 + \dots + Q_n)^{-\sigma} \alpha_n \left(1 - \frac{\alpha_{n-1}}{\alpha_n}\left(1 - \frac{Q_n}{Q_1 + \dots + Q_n}\right)^{-\sigma}\right) \\
            &= (Q_1 + \dots + Q_n)^{-\sigma} \alpha_n \left((\rho - \sigma)\frac{Q_n}{Q_1 + \dots + Q_n} + o\left(\frac{Q_n}{Q_1 + \dots + Q_n}\right)\right) as \enspace n \rightarrow +\infty \\
            &\sim_{+\infty} (Q_1 + \dots + Q_n)^{-\sigma} \alpha_n (\rho - \sigma) \frac{Q_n}{Q_1 + \dots + Q_n}
        \end{align*}
        which implies that $\left((Q_1 + \dots + Q_n)^{-\sigma} \alpha_n\right)_{n \geq 1}$ is eventually increasing. Similarly $\left((Q_1 + \dots + Q_n)^{-\tau} \alpha_n\right)_{n \geq 1}$ is eventually decreasing.
        \item $(iv) \implies (iii)$ Set $\rho_n := \displaystyle\frac{Q_1 + \dots + Q_n}{Q_n}\left(1 - \displaystyle\frac{\alpha_{n-1}}{\alpha_n}\right), n \geq 2$. There exists $N \geq 2$ such that for all $n \geq N$, 
        $$ (Q_1 + \dots + Q_{n-1})^{-\sigma}\alpha_{n-1} \leq (Q_1 + \dots + Q_n)^{-\sigma}\alpha_n $$
        and 
        $$ (Q_1 + \dots + Q_n)^{-\tau}\alpha_n \leq (Q_1 + \dots + Q_{n-1})^{-\tau}\alpha_{n-1}. $$
        It follows that
        $$ 1 - \left(1 - \frac{Q_n}{Q_1 + \dots + Q_n}\right)^{\sigma} \leq 1 - \frac{\alpha_{n-1}}{\alpha_n} \leq 1 - \left(1 - \frac{Q_n}{Q_1 + \dots + Q_n}\right)^{\tau} $$
        which leads to $\sigma + o(1) \leq \rho_n \leq \tau + o(1)$ as $n \rightarrow +\infty$. Thus for all $\sigma, \tau$ such that $\sigma < \rho < \tau$,
        $$ \sigma \leq \liminf\limits_{n \rightarrow +\infty}{\rho_n} \leq \limsup\limits_{n \rightarrow +\infty}{\rho_n} \leq \tau. $$
    \end{itemize}
    \item $(i) \iff (iii)$
    \begin{itemize}
        \item $(i) \implies (iii)$ By Lemma \ref{lemma 1} we already have $\lim\limits_{n \rightarrow +\infty}{\mathcal{R}_n} = \displaystyle\frac{1}{1 + \rho}$. Furthermore for every $n \geq 2$,
        \begin{equation}
        \label{recurrence relation proof theorem 2}
            \mathcal{R}_n = \left(1 - \frac{Q_n}{Q_1 + \dots + Q_n}\right)\frac{\alpha_{n-1}}{\alpha_n}\mathcal{R}_{n-1} + \frac{Q_n}{Q_1 + \dots + Q_n}
        \end{equation}
        and
        \begin{align*}
            \frac{\mathcal{R}_n}{\mathcal{R}_{n-1}} &= \left(1 - \frac{Q_n}{Q_1 + \dots + Q_n}\right)\left(1 - \rho\frac{Q_n}{Q_1 + \dots + Q_n} + o\left(\frac{Q_n}{Q_1 + \dots + Q_n}\right)\right) \\
            &\qquad + \frac{Q_n}{Q_1 + \dots + Q_n}\left(1 + \rho + o(1)\right) \\
            &= 1 + o\left(\frac{Q_n}{Q_1 + \dots + Q_n}\right)
        \end{align*}
        as $n \rightarrow +\infty$.
        \item $(iii) \implies (i)$ From (\ref{recurrence relation proof theorem 2}) we have for all $n \geq 2$,
        \begin{align*}
            \frac{\alpha_{n-1}}{\alpha_n} &= \left(1 - \frac{Q_n}{Q_1 + \dots + Q_n}\right)^{-1}\left(\frac{\mathcal{R}_n}{\mathcal{R}_{n-1}} - (1 + \rho)\frac{Q_n}{Q_1 + \dots + Q_n} + o\left(\frac{Q_n}{Q_1 + \dots + Q_n}\right)\right) \\
            &= 1 - \rho \frac{Q_n}{Q_1 + \dots + Q_n} + o\left(\frac{Q_n}{Q_1 + \dots + Q_n}\right)
        \end{align*}
        as $n \rightarrow +\infty$.
    \end{itemize}
    \item $(ii) \iff (iv)$
    \begin{itemize}
        \item $(ii) \implies (iv)$ By Lemma \ref{lemma 2} $(ii) \implies (iii)$ and we already know that $(iii) \implies (iv)$.
        \item $(iv) \implies (ii)$ By considering $\alpha_n (Q_1 + \dots + Q_n)^{-\rho}$ instead of $\alpha_n$ we may without loss of generality assume $\rho = 0$. Set $\beta_n := \log \alpha_n$, $n \geq 1$. Let $\varepsilon > 0$. The sequence $(\beta_n - \varepsilon \log(Q_1 + \dots + Q_n))_{n \geq 1}$ is eventually decreasing and $(\beta_n + \varepsilon \log(Q_1 + \dots + Q_n))_{n \geq 1}$ eventually increasing, thus there exists $N \geq 1$ such that for all $n > N$,
        $$ \beta_n - \varepsilon \log(Q_1 + \dots + Q_n) \leq \beta_{n-1} - \varepsilon \log(Q_1 + \dots + Q_{n-1}) $$
        and
        $$ \beta_{n-1} + \varepsilon \log(Q_1 + \dots + Q_{n-1}) \leq \beta_n + \varepsilon \log(Q_1 + \dots + Q_n), $$
        which gives
        $$ -\varepsilon \log\left(\frac{Q_1 + \dots + Q_n}{Q_1 + \dots + Q_{n-1}}\right) \leq \beta_n - \beta_{n-1} \leq \varepsilon \log\left(\frac{Q_1 + \dots + Q_n}{Q_1 + \dots + Q_{n-1}}\right) $$
        that can be written for all $n > N$,
        $$ |\beta_n - \beta_{n-1}| \leq \varepsilon \left|\log\left(1 - \frac{Q_n}{Q_1 + \dots + Q_n}\right)\right| $$
        and it follows that $\lim\limits_{n \rightarrow +\infty}{\displaystyle\frac{Q_1 + \dots + Q_n}{Q_n}(\beta_n - \beta_{n-1})} = 0$. Take $\beta_0 \in \mathbb{R}$ and set $\theta_n := \displaystyle\frac{Q_1 + \dots + Q_n}{Q_n}(\beta_n - \beta_{n-1})$, $n \geq 1$.  It follows that for every $n \in \mathbb{N}$,
        \begin{align*}
            \beta_n &= \beta_0 + \sum_{k=1}^{n}{(\beta_k - \beta_{k-1})} \\
            &= \beta_0 + \sum_{k=1}^{n}{\frac{Q_k}{Q_1 + \dots + Q_k}\theta_k} \\
            &= \beta_0 + \sum_{k=1}^{n}{\int_{Q_1 + \dots + Q_{k-1}}^{Q_1 + \dots + Q_k}{\frac{e(x)}{[x]_Q}\,\mathrm{d}x}}
        \end{align*}
        where for every $x \geq 0$, $e(x) := \displaystyle\sum_{n=1}^{+\infty}{\theta_n \mathbbm{1}_{\{Q_1 + \dots + Q_{n-1} < x \leq Q_1 + \dots + Q_n\}}}$. Whence for all $n \in \mathbb{N}$,
        \begin{align*}
            \beta_n &= \beta_0 + \int_{0}^{Q_1 + \dots + Q_n}{\frac{e(x)}{[x]_Q}\,\mathrm{d}x} \\
            &= \beta_0 + \int_{0}^{Q_1 + \dots + Q_n}{\frac{\theta(x)}{x}\,\mathrm{d}x}
        \end{align*}
        with $[x]_Q := \min\left\{y \in Q[\mathbb{N^*}] \,|\, x \leq y\right\}$ and $\theta(x) := \displaystyle\frac{e(x)x}{[x]_Q} \xrightarrow[x \rightarrow +\infty]{} 0$ where $Q := (Q_n)_{n \geq 1}$ and $Q[\mathbb{N}^*] := \left\{\displaystyle\sum_{k=1}^{n}{Q_k}\,\bigg| n \geq 1\right\}$.
    \end{itemize}
\end{itemize}
\end{proof}

\subsection{Proof of Theorem \ref{theorem equilibrium II}}

We proceed in four steps by proving $(i) \implies (ii)$, $(ii) \implies (iii)$, $(iii) \implies (i)$ and then $(i) \iff (iv)$.

\begin{proof}
\begin{itemize}
    \item $(i) \implies (ii)$ Set for every $x \geq 0$,
    \begin{align*}
        f(x) &:= \sum_{n=1}^{+\infty}{\mathbb{E}\left[\varepsilon \left(S_{\tau_n} - S_{\tau_1^-}\right)\bigg|\mathcal{F}_n\right] \mathbbm{1}_{\{Q_1 + \dots + Q_{n-1} < x \leq Q_1 + \dots + Q_n\}}} \\
        n_Q(x) &:= \min\left\{n \in \mathbb{N}^*\,\bigg|\, x \leq \displaystyle\sum_{k=1}^{n}{Q_k}\right\} \\
        [x]_Q &:= \min\left\{y \in Q[\mathbb{N^*}] \,|\, x \leq y\right\} \numberthis \label{Q ceil function}
    \end{align*}
    where $Q := (Q_n)_{n \geq 1}$ and $Q[\mathbb{N}^*] := \left\{\displaystyle\sum_{k=1}^{n}{Q_k}\,\bigg| n \geq 1\right\}$. By convention we will consider that a sum with no terms will evaluate to $0$. Hence $f$ is positive, locally bounded and for every $x > 0$,
    $$ \frac{1}{x f(x)}\int_{0}^{x}{f(t)\,\mathrm{d}t} = \frac{[x]_Q}{x}\mathcal{R}_{n_Q(x)} - \left(\frac{[x]_Q}{x} - 1\right). $$
    Since $[x]_Q \sim_{+\infty} x$ and $\lim\limits_{x \rightarrow +\infty}{n_Q(x)} = +\infty$, we have
    $$ \lim\limits_{x \rightarrow +\infty}{\frac{1}{x f(x)}\int_{0}^{x}{f(t)\,\mathrm{d}t}} = \frac{1}{1 + \rho}, $$
    hence $f$ varies regularly with index $\rho$ (Theorem 1.6.1 in \cite{bingham_goldie_teugels_1987}) and may be written in the form
    $$ f(x) =  x^{\rho} \exp\left(\eta(x) + \int_{0}^{x}{\frac{\theta(t)}{t}\,\mathrm{d}t}\right),\quad x \geq 0 $$
    where $\eta$ and $\theta$ are two measurable functions such that $\lim\limits_{x \rightarrow +\infty}{\eta(x)} = \kappa \in \mathbb{R}$ and $\lim\limits_{x \rightarrow +\infty}{\theta(x)} = 0$ (Theorem 1.3.1 in \cite{bingham_goldie_teugels_1987}). Finally for every $n \geq 1$, 
    \begin{align*}
        \mathbb{E}\left[\varepsilon \left(S_{\tau_n} - S_{\tau_1^-}\right)\bigg|\mathcal{F}_n\right] &= f(Q_1 + \dots + Q_n) \\
        &= (Q_1 + \dots + Q_n)^{\rho}\exp\left(\eta(Q_1 + \dots + Q_n) + \int_{0}^{Q_1 + \dots + Q_n}{\frac{\theta(u)}{u}\,\mathrm{d}u}\right).
    \end{align*}
    \item $(ii) \implies (iii)$
    $$ \mathbb{E}\left[\varepsilon \left(S_{\tau_n} - S_{\tau_1^-}\right)\bigg|\mathcal{F}_n\right] \sim_{+\infty} \lambda (Q_1 + \dots + Q_n)^{\rho}\exp\left(\int_{0}^{Q_1 + \dots + Q_n}{\frac{\theta(u)}{u}\,\mathrm{d}u}\right), \quad \lambda > 0. $$
    Set $\alpha_n := \lambda (Q_1 + \dots + Q_n)^{\rho}\exp\left(\displaystyle\int_{0}^{Q_1 + \dots + Q_n}{\frac{\theta(u)}{u}\,\mathrm{d}u}\right)$, $n \geq 1$. By Lemma \ref{lemma 2} we have
    $$ \frac{\alpha_{n-1}}{\alpha_n} = 1 - \rho\frac{Q_n}{Q_1 + \dots + Q_n} + o\left(\frac{Q_n}{Q_1 + \dots + Q_n}\right) $$
    as $n \rightarrow +\infty$.
    \item $(iii) \implies (i)$ If $\rho > 0$, then
    \begin{align*}
        \log\left(\frac{\alpha_n}{\alpha_{n-1}}\right) &= \rho \frac{Q_n}{Q_1 + \dots + Q_n} + o\left(\frac{Q_n}{Q_1 + \dots + Q_n}\right) \enspace \text{as} \enspace n \rightarrow +\infty \\
        &\sim_{+\infty} \rho \frac{Q_n}{Q_1 + \dots + Q_n},
    \end{align*}
    which implies that $\lim\limits_{n \rightarrow +\infty}{\alpha_n} = +\infty$, whence $\displaystyle\sum_{n=1}^{+\infty}{Q_n \alpha_n} = +\infty$ and 
    $$ \sum_{k=1}^{n}{Q_k\, \mathbb{E}\left[\varepsilon\left(S_{\tau_k} - S_{\tau_1^-}\right)\bigg|\mathcal{F}_k\right]} \sim_{+\infty} \sum_{k=1}^{n}{Q_k \alpha_k}. $$
    This implies that
    $$ \mathcal{R}_n \sim_{+\infty} \frac{\displaystyle\sum_{k=1}^{n}{Q_k \alpha_k}}{\left(\displaystyle\sum_{k=1}^{n}{Q_k}\right) \alpha_n} \enspace \text{and} \lim\limits_{n \rightarrow +\infty}{\mathcal{R}_n} = \frac{1}{1 + \rho} $$
    by Lemma \ref{lemma 1}. Now let us take $\rho = 0$ and set $\theta > 0$ and $(\alpha'_n)_{n \geq 1}$ such that for every $n \geq 1$, $\alpha'_n := (Q_1 + \dots + Q_n)^{\theta} \alpha_n$. We have
    $$ \frac{\alpha'_{n-1}}{\alpha'_n} = 1 - \theta \frac{Q_n}{Q_1 + \dots + Q_n} + o\left(\frac{Q_n}{Q_1 + \dots + Q_n}\right) \enspace \text{as} \enspace n \rightarrow +\infty $$
    which implies by Lemma \ref{lemma 1}
    $$ \lim\limits_{n \rightarrow +\infty}{\frac{\displaystyle\sum_{k=1}^{n}{Q_k \alpha'_k}}{\left(\displaystyle\sum_{k=1}^{n}{Q_k}\right) \alpha'_n}} = \frac{1}{1 + \theta} $$
    for each $\theta > 0$. By setting $\mathcal{S}_n := Q_1 + \dots + Q_n$, $n \geq 1$ we have
    $$ \frac{\displaystyle\sum_{k=1}^{n}{Q_k \alpha'_k}}{\left(\displaystyle\sum_{k=1}^{n}{Q_k}\right) \alpha'_n} = \frac{\displaystyle \sum_{k=1}^{n}{Q_k \left(\frac{\mathcal{S}_k}{\mathcal{S}_n}\right)^{\theta}\, \mathbb{E}\left[\varepsilon\left(S_{\tau_k} - S_{\tau_1^-}\right)\bigg|\mathcal{F}_k\right]}}{\left(\displaystyle\sum_{k=1}^{n}{Q_k}\right)\mathbb{E}\left[\varepsilon\left(S_{\tau_n} - S_{\tau_1^-}\right)\bigg|\mathcal{F}_n\right]} \leq \mathcal{R}_n $$
    which implies that for all $\theta > 0$, $\displaystyle\frac{1}{1 + \theta} \leq \liminf\limits_{n \rightarrow +\infty}{\mathcal{R}_n}$ and $\liminf\limits_{n \rightarrow +\infty}{\mathcal{R}_n} \geq 1$. Besides from Assumption \ref{assumption a.s. VWAP constraint} we already know that $\limsup\limits_{n \rightarrow +\infty}{\mathcal{R}_n} \leq 1$ which leads to the desired result.
    \item $(i) \iff (iv)$ Set $\mathcal{U}_n := \displaystyle\sum_{k=1}^{n}{Q_k\, \mathbb{E}\left[\varepsilon\left(S_{\tau_k} - S_{\tau_1^-}\right)\bigg|\mathcal{F}_k\right]}$, $n \geq 1$.
    \begin{itemize}
        \item $(i) \implies (iv)$ For every $n \geq 2$,
        \begin{equation}
        \label{U_n condition theorem 1} 
            \frac{Q_1 + \dots + Q_n}{Q_n}\left(1 - \frac{\mathcal{U}_{n-1}}{\mathcal{U}_{n}}\right) = \frac{1}{\mathcal{R}_n}.
        \end{equation}
        It follows that
        $$ \lim\limits_{n \rightarrow +\infty}{\frac{Q_1 + \dots + Q_n}{Q_n}\left(1 - \frac{\mathcal{U}_{n-1}}{\mathcal{U}_{n}}\right)} = 1 + \rho $$
        and from Theorem \ref{theorem equilibrium I} there exist $\zeta \in \mathbb{R}$ and $\chi$ a bounded measurable function of a real variable such that for all $n \geq 1$,
        \begin{equation}
        \label{U_n}
            \mathcal{U}_n = (Q_1 + \dots + Q_n)^{1 + \rho}\exp\left(\zeta + \int_{0}^{Q_1 + \dots + Q_n}{\frac{\chi(u)}{u}\,\mathrm{d}u}\right).
        \end{equation}
        \item $(iv) \implies (i)$ Since we have (\ref{U_n})
        $$ \lim\limits_{n \rightarrow +\infty}{\frac{Q_1 + \dots + Q_n}{Q_n}\left(1 - \frac{\mathcal{U}_{n-1}}{\mathcal{U}_{n}}\right)} = 1 + \rho $$
        from Lemma \ref{lemma 2}. Then from (\ref{U_n condition theorem 1}) this implies $\lim\limits_{n \rightarrow +\infty}{\mathcal{R}_n} = \displaystyle\frac{1}{1 + \rho}$.
    \end{itemize}
\end{itemize}
\end{proof}

\subsection{Proof of Proposition \ref{proposition rho infinity theorem equilibrium II}}

\begin{proof}
Set for all $n \in \mathbb{N}^*$, 
$$ \mathcal{Z}_n := \frac{\displaystyle\sum_{k=1}^{n}{Q_k \alpha_k}}{\left(\displaystyle\sum_{k=1}^{n}{Q_k}\right) \alpha_n}. $$
It follows $\lim\limits_{n \rightarrow +\infty}{\mathcal{Z}_n} = 0$ from Lemma \ref{lemma 1}. Furthermore since $(\alpha_n)_{n \geq 1}$ is eventually increasing and $\displaystyle\sum_{n=1}^{+\infty}{Q_n} = +\infty$, we have $\displaystyle\sum_{k=1}^{n}{Q_k \alpha_k} \sim_{+\infty} \displaystyle\sum_{k=1}^{n}{Q_k \mathbb{E}\left[\varepsilon\left(S_{\tau_k} - S_{\tau_1^-}\right)\bigg|\mathcal{F}_k\right]}$ and then $\mathcal{Z}_n \sim_{+\infty} \mathcal{R}_n$.
\end{proof}

\subsection{Proof of Corollary \ref{corollary slowly varying function theorem equilibrium II}}

\begin{proof}
For all $n \geq 1$, 
$$ \langle\mathcal{I} \rangle_n = (Q_1 + \dots + Q_n)^{\rho}\exp\left(\zeta + \int_{0}^{Q_1 + \dots + Q_n}{\frac{\chi(u)}{u}\,\mathrm{d}u}\right) $$
and
$$ \mathcal{I}_n =  (Q_1 + \dots + Q_n)^{\rho}\exp\left(\eta(Q_1 + \dots + Q_n) + \int_{0}^{Q_1 + \dots + Q_n}{\frac{\theta(u)}{u}\,\mathrm{d}u}\right). $$
Set 
$$ \ell(x) :=  \exp\left(\zeta - \eta(x) + \int_{0}^{x}{\frac{\chi(u) - \theta(u)}{u}\,\mathrm{d}u}\right), \enspace x \geq 0 $$
hence $\ell$ is a slowly varying function (Theorem 1.3.1 in \cite{bingham_goldie_teugels_1987}) and we have for every $n \geq 1$, $\mathcal{R}_n = \ell(Q_1 + \dots + Q_n)$.
\end{proof}

\subsection{Proof of Corollary \ref{corollary log equivalent theorem equilibrium II}}

\begin{proof}
Set $f(x) := x^{\rho} \exp\left(\eta(x) + \displaystyle\int_{0}^{x}{\frac{\theta(u)}{u}\mathrm{d}u}\right)$, $x \geq 0$ with $\eta$ and $\theta$  two bounded measurable functions of a real variable converging to a finite number and to zero as $x$ goes to infinity. For every $x > 0$,
$$ \log f(x) = \rho \log x + \underbrace{\eta(x) + \int_{0}^{x}{\frac{\theta(u)}{u}\,\mathrm{d}u}}_{o(\log x) \enspace\text{as}\enspace x \rightarrow +\infty}, $$
hence when $\rho > 0$, $\log f(x) \sim_{+\infty} \rho \,log x$. Since
$$ \mathbb{E}\left[\varepsilon \left(S_{\tau_n} - S_{\tau_1^-}\right)\bigg|\mathcal{F}_n\right] = f(Q_1 + \dots + Q_n) $$
for all $n \geq 1$ and $\lim\limits_{n \rightarrow +\infty}{\left(Q_1 + \dots + Q_n\right)} = +\infty$, it follows that
$$ \log\mathbb{E}\left[\varepsilon \left(S_{\tau_n} - S_{\tau_1^-}\right)\bigg|\mathcal{F}_n\right] \sim_{+\infty}  \rho \log\left(Q_1 + \dots + Q_n\right). $$ 
\end{proof}

\subsection{Proof of Corollary \ref{corollary square root theorem equilibrium II}}

\begin{proof}
Let $\eta$, $\theta$ two bounded measurable functions of a real variable such that for all $n \geq 1$,
$$ \mathbb{E}\left[\varepsilon \left(S_{\tau_n} - S_{\tau_1^-}\right)\bigg|\mathcal{F}_n\right] = f(Q_1 + \dots + Q_n) $$
where for every $x \geq 0$,
$$ f(x) = x^{\rho} \exp\left(\eta(x) + \int_{0}^{x}{\frac{\theta(u)}{u}\mathrm{d}u}\right), $$ $\lim\limits_{x \rightarrow +\infty}{\eta(x)} = \kappa \in \mathbb{R}$ and $\lim\limits_{x \rightarrow +\infty}{\theta(x)} = 0$. It follows that
$$ \lim\limits_{n \rightarrow +\infty}{\frac{\mathbb{E}\left[\varepsilon \left(S_{\tau_n} - S_{\tau_1^-}\right)\bigg|\mathcal{F}_n\right]}{\hat{\sigma}_n}} = \left(\frac{Q}{V}\right)^{\rho}. $$
\end{proof}

\subsection{Proof of Corollary \ref{corollary embedding concave/convex market impact function theorem equilibrium II}}

\begin{proof}
\begin{itemize}
    \item Let $X > 0$ such that $f$ is concave on $[X,+\infty)$. For all $x \geq X$,
    $$ \frac{f(x) + f(X)}{2}(x-X) \leq \int_{X}^{x}{f(t)\,\mathrm{d}t}, $$
    which implies
    \begin{equation}
    \label{equivalent karamata's theorem}
    \frac{1}{2} \leq \frac{\displaystyle\int_{X}^{x}{f(t)\,\mathrm{d}t}}{(f(x) + f(X))(x - X)} \sim_{+\infty} \frac{\displaystyle\int_{X}^{x}{f(t)\,\mathrm{d}t}}{x f(x)}
    \end{equation}
    since $\lim\limits_{x \rightarrow +\infty}{f(x)} = +\infty$. Besides as
    $$ \lim\limits_{x \rightarrow +\infty}{\frac{\displaystyle\int_{X}^{x}{f(t)\,\mathrm{d}t}}{x f(x)}} = \displaystyle\frac{1}{1 + \rho} $$
    (Theorem 1.5.11 in \cite{bingham_goldie_teugels_1987}), it follows that $\displaystyle\frac{1}{2} \leq \displaystyle\frac{1}{1 + \rho}$ and $\rho \leq 1$ from (\ref{equivalent karamata's theorem}).
    \item The same leads to $\rho \geq 1$ when $f$ is eventually convex. 
\end{itemize}
\end{proof}

\subsection{Proof of Theorem \ref{theorem set of limit points}}

\begin{proof}
Set $\alpha_n := \mathbb{E}\left[\varepsilon \left(S_{\tau_n} - S_{\tau_1^-}\right)\bigg|\mathcal{F}_n\right]$, $n \geq 1$. $\mathcal{L}$ being the set of the limit points of a bounded sequence is compact. Furthermore
$$ 0 \leq \liminf\limits_{n \rightarrow +\infty}{\mathcal{R}_n} \leq \limsup\limits_{n \rightarrow +\infty}{\mathcal{R}_n} \leq 1. $$
Now we want to show that $\mathcal{L}$ is an interval. Let $\alpha$, $\beta$ $\in \mathcal{L}$ such that $\alpha < \beta$. Let $\gamma \in (\alpha,\beta)$, $\varepsilon > 0$ and $N \geq 1$ such that $\varepsilon < \min(\frac{\gamma - \alpha}{2}, \frac{\beta - \gamma}{2})$ and $\left((Q_1 + \dots + Q_n)\alpha_n\right)_{n \geq N}$ non-decreasing.
\begin{align*}
    \mathcal{R}_{n+1} &= \frac{Q_1 + \dots + Q_n}{Q_1 + \dots + Q_{n+1}}\frac{\alpha_n}{\alpha_{n+1}}\mathcal{R}_n + \frac{Q_{n+1}}{Q_1 + \dots + Q_{n+1}} \\
    &\leq \mathcal{R}_n + \frac{Q_{n+1}}{Q_1 + \dots + Q_{n+1}}.
\end{align*}
Hence there exist $n_0 \geq N$ such that for all $n \geq n_0$, $\mathcal{R}_{n+1} - \mathcal{R}_n < \varepsilon$, $n_1 \geq n_0$ such that $\mathcal{R}_{n_1} \in \,(\alpha - \varepsilon, \alpha + \varepsilon)$ and $n_2 > n_1$ such that $\mathcal{R}_{n_2} \in \,(\beta - \varepsilon, \beta + \varepsilon)$. Let
$$ A := \big\{n \geq n_1 \,\big|\, \mathcal{R}_n > \gamma + \varepsilon \big\}. $$
Since $n_2 \in A$, $A$ is a non empty subset of $\mathbb{N}$ and thus has a least element $m$. Besides $m > n_1$ since $\mathcal{R}_{n_1} \leq \gamma + \varepsilon$. By definition $\mathcal{R}_{m-1} \leq \gamma + \varepsilon$ and $\mathcal{R}_{m-1} = \mathcal{R}_m - (\mathcal{R}_m - \mathcal{R}_{m-1}) > \gamma + \varepsilon - \varepsilon = \gamma$ with $m \geq N$. Hence $\gamma \in \mathcal{L}$ and $[\alpha, \beta] \subset \mathcal{L}$. Thus $\mathcal{L}$ is a compact interval given by
$$ \mathcal{L} = \left[\liminf\limits_{n \rightarrow +\infty}{\mathcal{R}_n}\,,\, \limsup\limits_{n \rightarrow +\infty}{\mathcal{R}_n}\right]. $$
\end{proof}

\subsection{Proof of Proposition \ref{proposition average normalized market impact function}}

\begin{proof}
$\mathbb{P}^{\mathscr{C}}-$a.s. the function $x \in (0,1] \mapsto x^{\rho}$ is continuous and for every $x \in (0,1]$, $x^{\rho} \leq 1$, hence $\psi$ is continuous. Similarly $\lim\limits_{x \rightarrow 0^+}{\psi(x)} = \mathbb{P}^{\mathscr{C}}(\rho = 0)$ since $\mathbb{P}^{\mathscr{C}}-$a.s. $\lim\limits_{x \rightarrow 0^+}{x^{\rho}} = \mathbbm{1}_{\{\rho = 0\}}$.
\begin{itemize}
    \item If $\rho \leq 1$ (resp. $\geq 1$) $\mathbb{P}^{\mathscr{C}}-$a.s., then $x \mapsto x^{\rho}$ is concave (resp. convex) $\mathbb{P}^{\mathscr{C}}-$a.s. and $\psi$ is concave (resp. convex).
    \item For every $x > 0$, $\displaystyle\frac{\psi(x)}{x} = \mathbb{E}^{\mathscr{C}}[x^{\rho - 1}]$. If $\rho > 1$ $\mathbb{P}^{\mathscr{C}}-$a.s., then for all $x \in (0,1]$, $x^{\rho - 1} \leq 1$ $\mathbb{P}^{\mathscr{C}}-$a.s. leading to $\lim\limits_{x \rightarrow 0^+}{\displaystyle\frac{\psi(x)}{x}} = 0$ by the Lebesgue's dominated convergence theorem.
    \item By Fatou's lemma 
    $$ \liminf\limits_{x \rightarrow 0^+}{\displaystyle\frac{\psi(x)}{x}} = \liminf\limits_{x \rightarrow 0^+}{\mathbb{E}^{\mathscr{C}}\left[x^{\rho - 1}\right]} \geq \mathbb{E}^{\mathscr{C}}\left[\liminf\limits_{x \rightarrow 0^+}{x^{\rho - 1}}\right] \geq \mathbb{E}^{\mathscr{C}}\left[\liminf\limits_{x \rightarrow 0^+}{x^{\rho - 1}\mathbbm{1}_{\{0 < \rho < 1\}}}\right] = +\infty, $$
    and it follows that $\lim\limits_{x \rightarrow 0^+}{\displaystyle\frac{\psi(x)}{x}} = +\infty$.
    \item $x \in (0,1] \mapsto x^{\rho}$ is $\mathbb{P}^{\mathscr{C}}-$a.s. differentiable and $\displaystyle\frac{\mathrm{d}x^{\rho}}{\mathrm{d}x} = \rho x^{\rho - 1}$. Let $a \in (0,1]$ and $x \in [a,1]$. $\mathbb{P}^{\mathscr{C}}-$a.s. $\rho x^{\rho - 1} \leq \rho \mathbbm{1}_{\{\rho \geq 1\}} + a^{-1} \mathbbm{1}_{\{\rho < 1\}}$, thus $\psi$ is differentiable on $[a,1]$ and the conclusion follows.
\end{itemize}
\end{proof}

\subsection{Proof of Proposition \ref{proposition asymptotic size distribution}}

\begin{proof}
Let $n \geq 1$.
\begin{align*}
    \mathbb{P}^{\mathscr{C}}\left(N \geq n+1\,|\, N \geq n\right) &= \frac{\mathbb{P}^{\mathscr{C}}\left(N \geq n+1\right)}{\mathbb{P}^{\mathscr{C}}\left(N \geq n\right)} \\
    &\sim_{+\infty} \frac{(n +1)^{-\beta}}{n^{-\beta}} \\
    &\sim_{+\infty} \left(1 + \frac{1}{n}\right)^{-\beta}.
\end{align*}
We have
$$ \mathbb{P}^{\mathscr{C}}\left(N = n\right) \leq \mathbb{P}^{\mathscr{C}}\left(n q_- \leq Q \leq n q_+\right) \leq \mathbb{P}^{\mathscr{C}}\left(n \leq N \leq \displaystyle\frac{q_+}{q_-} n \right) \leq \mathbb{P}^{\mathscr{C}}\left(N \geq n\right) $$
and
\begin{align*}
    \mathbb{P}^{\mathscr{C}}\left(N \geq n\right) &= \sum_{k \geq n}{\mathbb{P}^{\mathscr{C}}\left(N = k\right)} \\
    &\sim_{+\infty} \sum_{k \geq n}{\frac{C}{k^{1 + \beta}}} \\
    &\sim_{+\infty} C\int_{n}^{+\infty}{\frac{\mathrm{d}x}{x^{1 + \beta}}} \\
    &\sim_{+\infty} \frac{C}{\beta n^{\beta}}.
\end{align*}
Hence $\lim\limits_{n \rightarrow +\infty}{n^{1 + \beta} \mathbb{P}^{\mathscr{C}}\left(N = n\right)} = C > 0$ and $\lim\limits_{n \rightarrow +\infty}{n^{\beta} \mathbb{P}^{\mathscr{C}}\left(N \geq n\right)} = \displaystyle\frac{C}{\beta} > 0$ and we can take $C' := \displaystyle\frac{C}{2}$ and $C'' := \displaystyle\frac{2C}{\beta}$.
\end{proof}

\subsection{Proof of Proposition \ref{proposition convergence of the moments of the metaorder size}}

\begin{proof}
Set $\Delta_n := \mathbb{P}^{\mathscr{C}}(N=n) \mathbb{E}^{\mathscr{C}}\left[Q^{\nu}\,|\,N=n\right]$, $n \geq 1$.
\begin{itemize}
    \item $Q_{|N} \sim \mathcal{U}\left(\left[N q_-, N q_- + (q_+ - q_-)N^{1 - \gamma}\right]\right)$
    \begin{align*}
        \mathbb{E}^{\mathscr{C}}\left[Q^{\nu}\right] &= \sum_{n=1}^{+\infty}{\mathbb{P}^{\mathscr{C}}(N=n) \mathbb{E}^{\mathscr{C}}\left[Q^{\nu}\,|\,N=n\right]} \\
        &= \sum_{n=1}^{+\infty}{\mathbb{P}^{\mathscr{C}}(N=n) \frac{\left[n q_- + (q_+ - q_-)n^{1-\gamma}\right]^{1 + \nu} - (n q_-)^{1 + \nu}}{(1 + \nu)(q_+ - q_-)n^{1 - \gamma}}}.
    \end{align*}
    Hence
    $$ \Delta_n = \mathbb{P}^{\mathscr{C}}(N=n) \frac{\left[n q_- + (q_+ - q_-)n^{1-\gamma}\right]^{1 + \nu} - (n q_-)^{1 + \nu}}{(1 + \nu)(q_+ - q_-)n^{1 - \gamma}}, \enspace n \geq 1. $$
    If $\gamma = 0$, then
    $$ \Delta_n \sim_{+\infty} \frac{C( q_{+}^{1 + \nu} - q_{-}^{1 + \nu})}{(1 + \nu)(q_+ - q_-)} \frac{1}{n^{1 + \beta - \nu}}. $$
    Else $\gamma > 0$, and it follows that for all $n \geq 1$,
    \begin{align*}
        \Delta_n &= \mathbb{P}^{\mathscr{C}}(N=n) \frac{(n q_-)^{1 + \nu}\left(1 + \displaystyle\frac{q_+ - q_-}{q_- n^{\gamma}}\right)^{1 + \nu} - (n q_-)^{1 + \nu}}{(1 + \nu)(q_+ - q_-)n^{1 - \gamma}} \\
        &= \mathbb{P}^{\mathscr{C}}(N=n) \frac{q_{-}^{1 + \nu} n^{1 + \nu}\left[\left(1 + \displaystyle\frac{q_+ - q_-}{q_- n^{\gamma}}\right)^{1 + \nu} - 1\right]}{(1 + \nu)(q_+ - q_-)n^{1 - \gamma}},
    \end{align*}
    which gives
    \begin{align*}
        \Delta_n &\sim_{+\infty} \frac{C}{n^{1+\beta}}\frac{q_{-}^{\nu} n^{1 + \nu - \gamma} (1 + \nu)(q_+ - q_-)}{(1 + \nu)(q_+ - q_-)n^{1-\gamma}} \\
        &\sim_{+\infty} \frac{C q_{-}^{\nu}}{n^{1 + \beta - \nu}}.
    \end{align*}
    In the two cases we can conclude.
    \item $Q_{|N} \sim \mathcal{U}\left(\left[N q_+ - (q_+ - q_-)N^{1 - \gamma}, N q_+\right]\right)$. In that case, when $\gamma > 0$,
    $$ \Delta_n \sim_{+\infty} \frac{C q_{+}^{\nu}}{n^{1 + \beta - \nu}} $$
    and the same conclusion holds.
\end{itemize}
\end{proof}

\subsection{Proof of Proposition \ref{real metaorders and non-decreasing concave market impact functions}}

\begin{proof}
Set $\mathcal{S}_n := \displaystyle\sum_{k=1}^{n}{Q_k}$, $n \in \mathbb{N}$. For each $k \geq 1$, $\displaystyle\int_{\mathcal{S}_{k-1}}^{\mathcal{S}_k}{f(x)\,\mathrm{d}x} \leq \displaystyle\int_{\mathcal{S}_{k-1}}^{\mathcal{S}_k}{f(\mathcal{S}_k)\,\mathrm{d}x}$. Hence for all $n \geq 1$, $\displaystyle\int_{0}^{\mathcal{S}_n}{f(x)\,\mathrm{d}x} \leq \sum_{k=1}^{n}{Q_k f(\mathcal{S}_k)}$ and $\mathcal{R}_n \geq \displaystyle\frac{1}{\mathcal{S}_n f(\mathcal{S}_n)}\displaystyle\int_{0}^{\mathcal{S}_n}{f(x)\,\mathrm{d}x}$. Furthermore $f$ is concave and $f(0) = 0$, it follows that for every $x \geq 0$, $\displaystyle\frac{x f(x)}{2} \leq \displaystyle\int_{0}^{x}{f(t)\,\mathrm{d}t}$. Whence for all $n \geq 1$, $\mathcal{R}_n \geq \displaystyle\frac{1}{2}$ which gives $\mathcal{R}_N \in [1/2, 1]$ $\mathbb{P}^{\mathscr{C}}-$a.s.
\end{proof}

\subsection{Proof of Proposition \ref{proposition metaorder duration inequalities}}

\begin{proof}
$G$ is a decreasing homeomorphism onto its image, hence $G^{-1}$ is also decreasing and $G^{-1}(1) \leq T_N \leq G^{-1}\left(\displaystyle\frac{1}{2}\right)$ since $\displaystyle\frac{1}{2} \leq \mathcal{R}_N \leq 1$ and $G^{-1}(1) = 0$. Furthermore if $G$ is convex we have $G\left(\mathbb{E}^{\mathscr{C}}\left[T_N\right]\right) \leq \mathbb{E}^{\mathscr{C}}\left[G\left(T_N\right)\right]$ which gives $\mathbb{E}^{\mathscr{C}}\left[T_N\right] \leq G^{-1}\left(\mathbb{E}^{\mathscr{C}}\left[\mathcal{R}_N\right]\right)$ since $G(T_N) = \mathcal{R}_N$.
\end{proof}

\newpage
\bibliographystyle{apalike}
\bibliography{bibliography}

\end{document}